\documentclass[showpacs,twocolumn,amssymb,preprintnumbers,prc,floatfix,amsmath]{revtex4-1}
\usepackage{rotating,epsfig,dcolumn}
\usepackage{graphicx,bm}
\usepackage{bm,color}
\usepackage{epstopdf}
\newcommand{\ef}{$e$fm$^2$\,}
\newcommand{\eff}{$e^2$fm$^4$\,}
\newcommand{\bet}{$B(E2:2^+\to 0^+)$\,}
\newcommand{\bL}{\begin{Large}}
\newcommand{\eL}{\end{Large}}
\newcommand{\bl}{\begin{large}}
\newcommand{\el}{\end{large}}
\newcommand{\be}{\begin{equation}}  
\newcommand{\ee}{\end{equation}}
\newcommand{\ba}{\begin{eqnarray*}}
\newcommand{\ea}{\end{eqnarray*}}

\newcommand{\hw}{$\hbar \omega \,$}

\newcommand{\ie}{{\it i.e.,\ }}
\newcommand{\q}{$\langle 2q^{20}\rangle\,$}
\newcommand{\jun}{{\sc jun45}~\cite{jun45}}

\def\tcn{\textcolor{black}}

\begin{document}

\title{Nilsson-SU3 selfconsistency in heavy N=Z nuclei}
\author{A.~P.~Zuker$^1$, A.~Poves$^2$, F.~Nowacki$^1$ and
  S.~M.~Lenzi$^3$} \affiliation{(1) Universit\'e de Strasbourg, IPHC,
  CNRS, UMR7178, 23 rue du Loess 67037 Strasbourg, France\\(2)
  Departamento de F\'isica Te\'orica e IFT-UAM/CSIC, \mbox{Universidad
  Aut\'onoma de Madrid, 28049 Madrid, Spain and} ISOLDE, CERN,
  CH-1211, Gen\`eve Suisse\\ (3) Dipartimento di Fisica e Astronomia
  dell'Universit\`a and INFN, Sezione di Padova, I-35131 Padova,
  Italy}

\begin{abstract}
It is argued that there exist natural shell model spaces optimally
adapted to the operation of two variants of Elliott's SU3 symmetry that
provide accurate predictions of quadrupole moments of deformed states.
A selfconsistent Nilsson-like calculation describes the competition
between the realistic quadrupole force and the central field,
indicating a {\em remarkable stability of the quadrupole
  moments}---which remain close to their quasi and pseudo SU3
values---as the single particle splittings increase.  A detailed study
of the $N=Z$ even nuclei from $^{56}$Ni to $^{96}$Cd reveals that the
region of prolate deformation is bounded by a pair of transitional
nuclei $^{72}$Kr and $^{84}$Mo in which prolate ground state bands are
predicted to dominate, though coexisting with oblate ones.
\end{abstract}
\date{\today}
\maketitle
\section{Introduction}
 Large Scale Shell Model calculations (LSSM), when doable, are the
 spectroscopic tool of choice in theoretical nuclear structure. When
 they are not doable it is often advised to \tcn{switch to}
 other---basically mean field---methods. A common feature of these
 approaches is the reliance on quadrupole degrees of freedom as the
 backbone of nuclear structure, which in shell model language
 translates as dominance of the quadrupole force, which is indeed (or
 should be) a classic view. Our task is to find ways to put to good
 use this dominance. It starts by discovering which are the model
 spaces in which to operate. The choice turns out to be quite unique
 (the EEI spaces to be defined soon). Though most often it leads to
 intractably large diagonalizations, it also happens to be tailored to
 take full advantage of two variants---pseudo and quasi-SU3---of
 Elliott's SU3 symmetry~\cite{su3}. After explaining in detail how
 these symmetries operate we turn to quantitative estimates of their
 reliability by defining and implementing a selfconsistent
 Nilsson~\cite{nilsson} approach in which the interplay of a realistic
 quadrupole interaction with the spherical central field establishes
 the resilience of the predicted quadrupole moments. The controlling
 parameters are the quadrupole moments themselves which in the absence
 of a central field reduce to one of their SU3-like guises.

These ideas are applied to the heavy even $N=Z$ nuclei shedding light
on the hitherto poorly understood competition between prolate and
oblate quadrupole coherence. In this region the full interplay of
quasi and pseudo SU3 schemes operates, illustrating what will become
the rule for well deformed nuclei---so far only schematically explored
at the onset of rotational motion at $N=90$~\cite{zrpc}.

\section{The natural ZBM (or EEI)  model spaces}\label{sec:eei}
The usual lore about shell model spaces is that for light and medium
nuclei they involve one major oscillator (HO) shell bounded by magic
numbers at $N,Z$=4, 8, 20 and 40 while for heavier systems the
spin-orbit (SO) force takes over and the magic boundaries move to
$N,Z$= 28, 50, 82 and 126. This view has some merit but misses two
crucial points: a) the observed shell evolution is not driven by the
SO terms present in the NN interactions, but by three body forces (a
word on this later); b) the correct model spaces are larger than those
defined by the SO boundaries. Let us examine the possible examples.

In the $p$ shell starting at $^{4}$He, as particles are added the
largest orbit $p_{3/2}$ is ``Extruded'' (or Ejected or Expelled) from
the space by becoming a ``closed shell'' when filled, while the
largest orbit in the next shell ``Intrudes'' so as to define the first
of the EI spaces $p_{1/2}d_{5/2}\equiv r_1d$ (closing at $^{28}$Si).
The notation $r_p$ stands for ``rest of the major shell of principal
quantum number $p$'' \ie all the orbits except the largest one. What
we miss here is that the $d_{5/2}$ intruder does not come alone but with
an $s_{1/2}$ partner, as made evident by the spectrum of
$^{13}$C~\cite{ensdf}. Therefore the correct space is the first of the 
Extended EI spaces: $r_1ds$ (EEI1 or ZBM~\cite{zbm}), with  $ds=d_{5/2}s_{1/2}$;
which is the first instance of a ``$\Delta j=2$'' sequence. 

\noindent
\tcn{{\sf Notation}.} The full harmonic oscillator shells are called $sd,\,
pf,\, sdg$\ldots while the reverse order $ds,\, fp,\, gds$\ldots will
be used for the $\Delta j=2$ sequences.

Next candidate comes from the $sd$ shell starting at $^{16}$O where,
as it fills, $d_{5/2}$ is separated from its partners while drawing
down the largest orbit in the next shell so as to define the EI2 space:
\mbox{$s_{1/2}d_{3/2}f_{7/2}\equiv r_2f$} (starting at $^{28}$Si and closing
at $^{56}$Ni). Except that we miss again that the intruder comes with
its $\Delta j=2$ partner (as seen in $^{29}$Si~\cite{ensdf}) so
$r_2f$ becomes $r_2fp$ (EEI2 \tcn{or ZBM2}) with $fp=f_{7/2}p_{3/2}$.
\tcn{Then} we find the space, relevant for this study,
\mbox{$p_{1/2}p_{3/2}f_{5/2}g_{9/2}\equiv r_3g$} (EI3 closing at $^{100}$Sn) 
which is expected to become $r_3gds$ (EEI3 \tcn{or ZBM3}) with
\mbox{$gds=g_{9/2}d_{5/2}s_{1/2}$}. Direct experimental evidence of the
presence of the $\Delta j=2$ partners is hard to obtain in this
region, but abundant indirect evidence will be presented in this
paper. 

\begin{figure}[h]
\begin{center}
  \includegraphics[width=1.0\columnwidth]{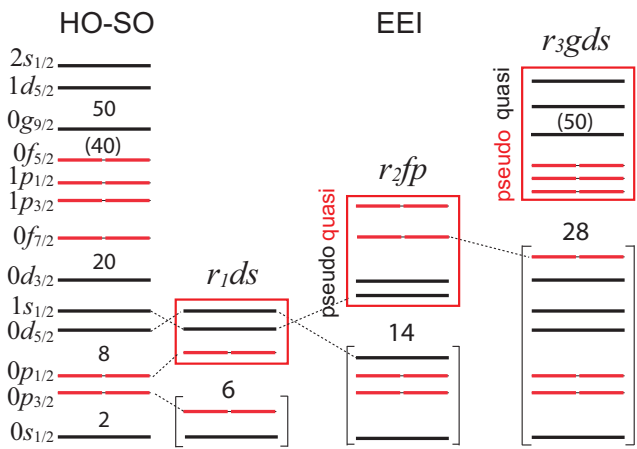}
\end{center}
\caption{\label{fig:EEI} (color online) Evolution of model spaces from
  Spin-orbit (SO) (around HO closures) to Extended Extruder-Intruder
  (EEI) made of Pseudo-SU3 and Quasi-SU3 subspaces (explained in
  Section~\ref{sec:qpsu3}).}
\end{figure}

\noindent
\tcn{{\sf Digression on shell formation.}} One objection to the
description above is that $^{12}$C and $^{28}$Si are not closed shells
(though $^{56}$Ni is, to a good approximation). However EI numbers at
$N,Z$=6, 14, 28, 50, 82 and 126 provide good boundaries and many
convincing candidates to magicity in the light nuclei (such as
$^{14}$C, $^{22}$O and $^{34}$Si) and the {\em only} systematic magic
numbers beyond. The transition from HO to EI major closures demands
three-body mechanisms whose irrefutable need is now established on
theoretic~\cite{abin3b} and empiric~\cite{zu3b, dz10} grounds.
Explicit introduction of three-body forces~\cite{ot3b} does not lead
so far to consistent agreement with the empirical results. The (hard
to sell) notation EI instead of the usual SO is meant to stress that
the spin orbit force---in the classic $l\cdot s$ sense---is perfectly
given by existing NN interactions above HO closures where it is
responsible for the largest orbit coming lowest~\cite{sz}. However, it
is definitely not responsible for the EI closures which demand
splittings much larger that the $l\cdot s$ one provided by the NN
interactions. To fix ideas: in $^{48}$Ca they would produce a
$f_{7/2}-p_{3/2}$ single particle gap equal to that in $^{41}$Ca \ie
2.5 MeV smaller than the observed one. A discrepancy that increases to
some 4.5 MeV in $^{56}$Ni.  \tcn{The evolution of subshell SO ordering
  on top of HO closures to the EEI patterns is illustrated in
  Fig.~\ref{fig:EEI} for different model spaces.}

Both $r_1ds$ (ZBM) and $r_2fp$ (\tcn{ZBM2} or SDFP) models lead to
feasible and successful diagonalizations in the neighborhood of
$^{16}$O and $^{40}$Ca~\cite{zbm,rmp}. The $r_3gds$ space is expected
to work equally well around $^{80}$Zr---formally the magic upper
boundary of the $pf$ shell---which turns out to be a splendid
rotor~\cite{lister}. A pure $pf$ description starts failing around $N,
Z\approx 34$, and it could be hoped that $r_3g$ would cope beyond, but
the calculations (always feasible though sometimes hard) fail to
produce strongly deformed prolate bands demanded by the data. Which
are naturally explained in the $r_3gds$ space as we shall demonstrate
notwithstanding the near impossibility of exact diagonalizations:
First through heuristic arguments based on the approximate SU3
symmetries, and then by very simple selfconsistent calculations that
account semi quantitatively for the interplay between the realistic
quadrupole interaction and the monopole central field.

 \section{Quadrupole coherence: SU3, pseudo-SU3 and  quasi-SU3}\label{sec:qpsu3}

Nuclear rotational motion was predicted by Bohr and Mottelson in
1953~\cite{bm53}. The idea was that nuclei could acquire a permanent
quadrupole deformation in their intrinsic frame, that would translate
into a $J(J+1)$ spectrum in the laboratory frame. Historically, this
first example of spontaneously broken symmetry was confronted with the
need to explain how a deformed intrinsic state---which has no definite
angular momentum $J$---could be an eigenstate of a system that must
necessarily conserve $J$. The elegant way out was found by Elliott
whose SU3 model~\cite{su3} provides a rigorous example of intrinsic
states that are not eigenstates of a Hamiltonian $H$ but of $H-\lambda
J(J+1)$.
\begin{figure}[h]
\begin{center}
  \includegraphics[width=0.9\columnwidth]{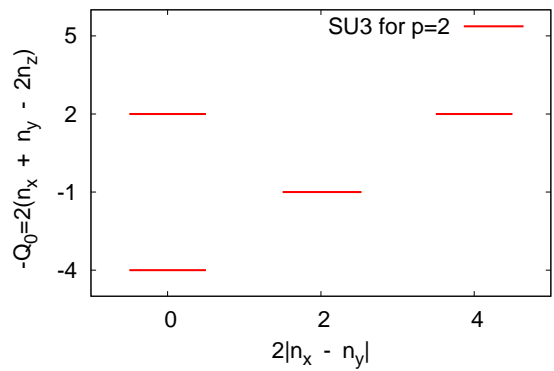}
\end{center}
\vspace*{-0.7cm}
\caption{\label{fig:SU3} (color online) Eigenstates of $-2q_{20}$.
  SU3 intrinsic states of minimum energy are obtained by orderly
  filling. Careful: we are plotting $-Q_0$.}
\end{figure}

More precisely, $H$ is taken to be the quadrupole force $-2q\cdot 2q$,
with $q\equiv q^{2m}=r^2C^{2m}=r^2 \sqrt{4\pi/5} \tcn{ \; Y^{2m}}$ acting in
a full major HO shell. Then the eigenstates have the form
$E(L,i)=E(i)+3 L(L+1)$, where $L$ is the orbital angular momentum and
$E(i)$ the energy of one of the possible intrinsic states. We shall be
interested only in those that maximize the intrinsic quadrupole moment
which we write in terms of oscillator quanta
$Q_0=2q^{20}=(2n_z-n_x-n_y)$. Taking for example $p=n_x+n_y+n_z=2$ the
six possible single particle states
$[n_zn_xn_y]$=[200],[110],[101],[020],[011],[002] can be disposed as
in Fig.~\ref{fig:SU3}.  The intrinsic states are the determinants
obtained by filling the fourfold degenerate orbits (two neutrons and
two protons of spins up and down) from below (prolate states with
$Q_0>0$) or from above (oblate states with $Q_0<0$). Prolate filling
is favored as it leads to larger $|Q_0|$.

Originally, SU3 was expected to apply to the $sd$ shell. And indeed,
the four particles in $^{20}$Ne ($Q_0=16$) produce a good rotor and
eight particles in $^{24}$Mg---because of the degeneracy of the
$Q_0=1$ levels in Fig.~\ref{fig:SU3}---lead to triaxiality, associated
to  \tcn{the mixing of} $K=0$ and $K=2$ prolate bands. For twelve particles in $^{28}$Si,
both shapes are expected to be degenerate ($|Q_0|=24$). Observation
does not quite square with predictions: the $K=2$ band in $^{24}$Mg is
higher than expected, and the ``nearly degenerate'' oblate and prolate
states in $^{28}$S are separated by some 6 MeV with a third candidate
coming in (the $d_{5/2}^{12}$ $N=Z=14$ closure in
Fig.~\ref{fig:EEI}). Still, the departure from strict SU3 validity
should not hide the fact that $^{24}$Mg has a $K=2$ ($\gamma$) band,
and that three of the six lowest states in $^{28}$Si have $J=0^+$, a
forerunner of other spectacular coexistence situations.
  
Though Elliott's conceptual breakthrough was obscured by the limited
applicability of the exact SU3 symmetry, its indicative value remains
high, as illustrated by examining the possible forms of the $q^{20}$
operator in $LS$ and $jj$ formalisms in Eqs.(1--5): They will be seen
to suggest naturally the pseudo and quasi SU3 variants that are the
backbone of a full shell model description of rotational motion.
\begin{gather}
\label{eq:q2}
\langle pl\vert r^2\vert pl\rangle =p+3/2\\ 
\langle pl\vert r^2\vert pl+2\rangle =-[(p-l)(p+l+3)]^{1/2} \\
\langle l m\vert C_2\vert l m\rangle
=\frac{l(l+1)-3m^2}{(2l+3)(2l-1)},~~~ \langle l m\vert C_2\vert l
+2m\rangle \nonumber \\
=\frac{3}{2}\left\{\frac
{[(l+2)^2-m^2][(l+1)^2-m^2]}{(2l+5)(2l+3)^2(2l+1)}\right\}^{1/2}\\
\langle jm\vert C_2\vert jm\rangle =\frac{j(j+1)-3m^2}{2j(2j+2)},~~~ 
\langle jm\vert C_2\vert j+2m\rangle \nonumber \\ 
 =\frac{3}{2} \left\{\frac{[(j+2)^2-m^2]
[(j+1)^2-m^2]}{(2j+2)^2(2j+4)^2}\right\}^{1/2}\\
\langle jm\vert C_2\vert j+1m\rangle =
-\frac{3m[(j+1)^2-m^2]^{1/2}}{j(2j+4)(2j+2)}\label{jj+1}
\end{gather}
Intrinsic states can be constructed by diagonalizing $q^{20}$ which
can be done in three possible ways, described after  \tcn{another} digression.

\noindent
 \tcn{{\sf Digression.}} So far we have assumed dimensionless oscillator
coordinates and made no difference between \q and $Q_0$. Dealing with
electromagnetic properties demands to recover dimensions so $r^2\to
r^2b^2$ where $b^2$ is the oscillator parameter. Then $Q_0\to
Q_0b^2$. On the other hand \q is best kept adimensional when working
with the quadrupole interaction.  So now $Q_0/b^2$=\q, and the choice
of notation will depend on context.

\subsubsection{Strict SU3}\label{sec:strict}

Use Eqs.~(1,2,3) in $LS$ form to obtain exactly Fig.~\ref{fig:SU3}.
Alternatively use Eqs.~(1,2,4,5) in $jj$ form to incorporate spin,
leading to the lower panel of Fig.~\ref{fig:su3}.  Only positive
values of $K\equiv |m|$ are shown. Each orbit may contain 2 neutrons
and 2 protons. Note that if in Fig.~\ref{fig:SU3} spin is allowed each
orbit splits into $2(n_x-n_y)\equiv 2m\to 2(m\pm 1/2)$ and the one to
one correspondence with the lower panel of Fig.~\ref{fig:su3} becomes
evident.

The importance of SU3 goes well beyond its mathematical elegance: it
rests on the introduction of the $q\cdot q$ interaction restricted to
a single major HO shell. Which, as demonstrated in~\cite{mdz}, is the
major collective ingredient of realistic Hamiltonians (\ie consistent
with two nucleon data).

\subsubsection{Pseudo SU3}\label{sec:pseudo}

Pseudo SU3~\cite{pseudo} is adapted to $r_p$ spaces whose orbits have
the same angular momentum $j$--sequences as those of full HO major
 \tcn{shell} with total quantum number $p-1$ and proceeds as if $r_p\equiv$
HO$(p-1)$, in our case $r_3\equiv sd$. For the angular Eqs.~(4,5) the
identity is perfect but the radial Eqs.~(1,2) raise a problem: $r_3$
has $p=3$ and $sd$ has $p=2$. The bottom panel of Fig.~\ref{fig:su3}
exhibits both the strict SU3 (or pseudo SU3) values for $p=$ 2 and 4,
as well as the exact result of diagonalizing $2q^{20}$ in the $r_3$
space, collected under {\sf p-d} in Table~\ref{tab:ex}. It is seen
that the differences are substantial but they do not invalidate the
existence of an underlying SU3 symmetry: the $q\cdot q$ interactions
in the $sd$ and $r_3$ spaces are very different but their behavior is
qualitatively similar. In what follows we always use the exact $r_3$
variant of $q\cdot q$.

\subsubsection{Quasi SU3}\label{sec:quasi}

\begin{figure}[h]
\begin{center}
  \includegraphics[width=1.0\columnwidth]{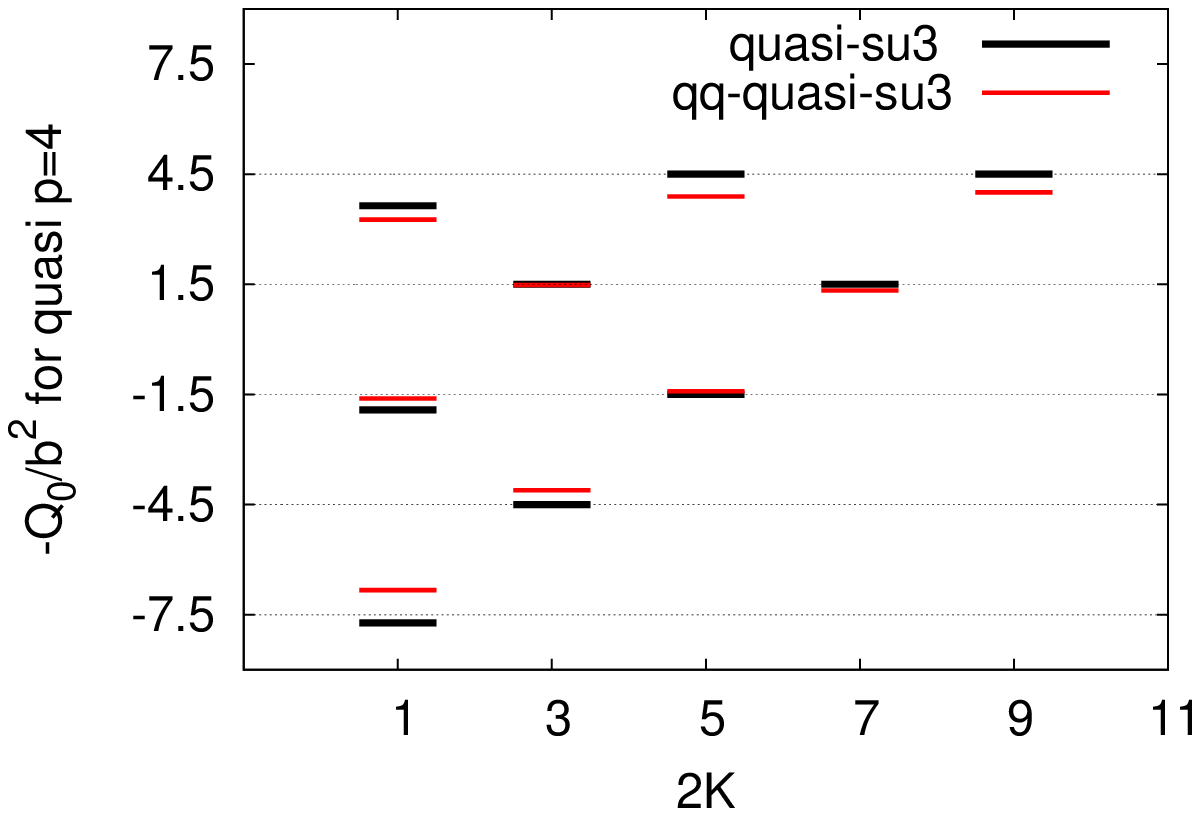}
  \includegraphics[width=1.0\columnwidth]{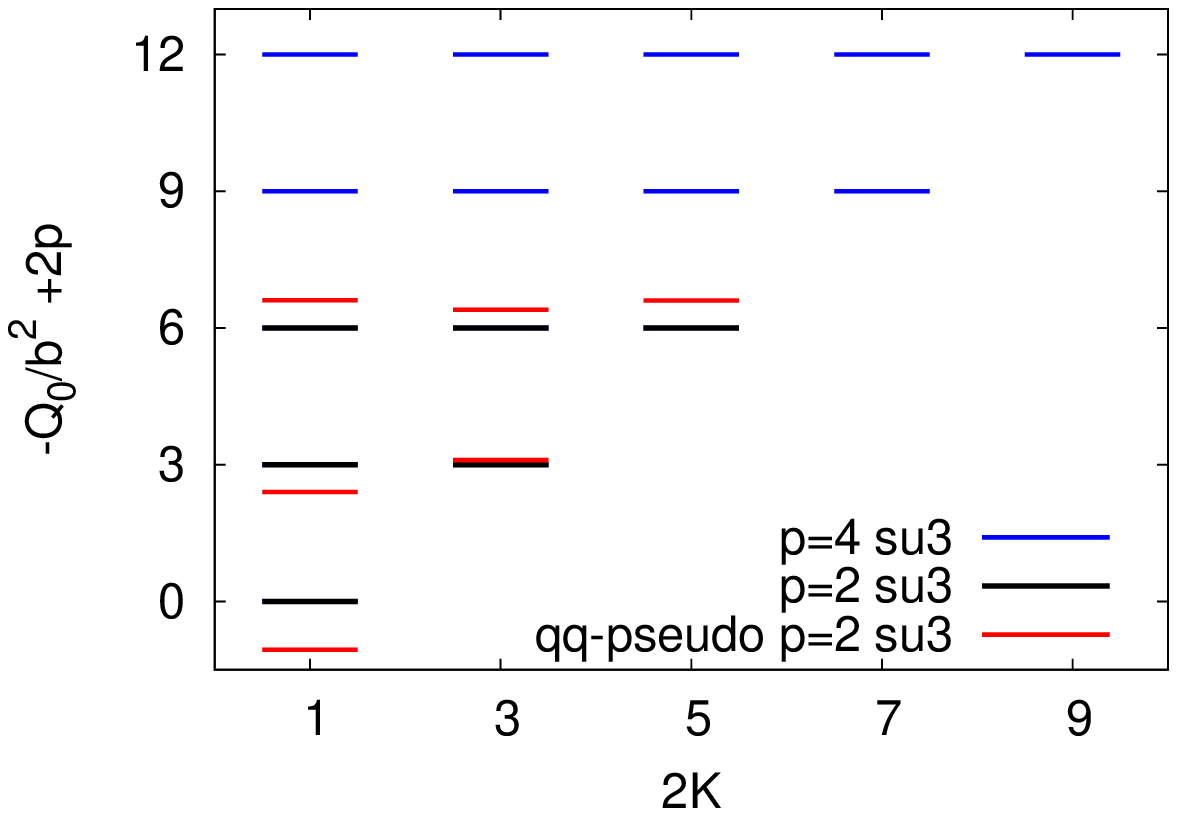}
\end{center}
\caption{\label{fig:su3} (color online) \tcn{{\bf The Zuker Retamosa
      Poves (ZRP) diagrams.}} Top: Intrinsic states in the $p$=4 $gds$
  space for the quasi-SU3 model and for the exact $q\cdot q$
  calculation. See text for explanation of thin lines. Bottom:
  Intrinsic states of SU3 or pseudo-SU3 for $p=2$ and 4. For the
  former exact $q\cdot q$ values ('qq-pseudo') are also shown. Each
  orbit may contain two neutrons and two protons. The lowest six
  orbits are common to $p=2$ and 4.}
\end{figure}

Quasi SU3~\cite{zrpc,a4749} is adapted to $\Delta j=2$ spaces. Then
$\langle jm\vert C_2\vert j+1m\rangle$ in Eq.~(\ref{jj+1}) plays no
role. Now identify the $\Delta j=2$ sequence to a $\Delta l=2$ one. In
our case $J=9/2,\,5/2,\,1/2$ to $l= 4,\,2,\,0$. Then replace Eqs.~(1,
2 and 4) by Eqs~(1, 2 and 3), through $l\to j$, $p\to p+1/2$, $m\to
m+1/2$ and $-m\to -m-1/2$: $(m>0)$. This defines a quasi-$q_{20}$
operator whose spectrum is shown (under 'quasi-su3') in the upper panel
of Fig.~\ref{fig:su3}, where thin lines indicate a one to one
correspondence with Fig.~\ref{fig:SU3} with bandheads at $2p-1/2$,
 except fo $K=1/2$ for even $p$. For odd $p$ the correspondence is
perfect throughout. The spectrum for the genuine  \tcn{$q^{20}$} operator
('qq-quasi-su3' in the figure) is seen to be quite close to the
schematic one.  (numerical values are collected under {\sf q-d} in
Table~\ref{tab:ex}).

\begin{table}[h]
\caption{\label{tab:ex} Eigenvalues of $-2q^{20}$ for the i-th
  quasi--$gds$ (denoted q) and pseudo--$r_3$ (denoted p) orbits; {\sf q-s}, {\sf p-s}
  for the results using the schematic quasi and pseudo quadrupole
  forces in Fig.~\ref{fig:su3}.  {\sf q-d}, {\sf p-d} are the results of
  diagonalizing the  \tcn{exact quadrupole} interaction; {\sf c-q} and {\sf c-p} are the
  corresponding cumulated absolute values for n particles in units of $b^2$}.
\begin{tabular*}{\linewidth}{@{\extracolsep{\fill}}|c|ccccccccc|}
\hline i& 1 &2 &3 &4 & 5 &6 &7 &8 &9\\ 
\hline
{\sf q-s}&-7.71&-4.50&-1.92&-1.50&1.50&1.50&3.64&4.50&4.50 \\ 
{\sf q-d}&-6.83&-4.11&-1.61&-1.42&1.33&1.48&3.26&3.90&4.00 \\ 
\hline
{\sf p-s}&-4.00&-1.00&-1.00& 2.00&2.00&2.00& & & \\ 
{\sf p-d}&-5.06&-1.41&-1.08& 2.37&2.57&2.61& & & \\
\hline
\multicolumn{10}{c}{}\\
\multicolumn{10}{c}{$Q_0$ values for $n$ particles ({\sf c-q} for $gds$ and {\sf c-p} for $r_3$)}\\
\hline
{\sf n}& 4 &8 &12 &16 & 20 &24 &28 &32 &36\\
\hline
{\sf c-q}&27.32&43.76&50.20&55.88&50.56&44.64&31.60&16.00&0.00 \\  
{\sf c-p}&20.24&25.88&30.20&20.72&10.44&0.00  & & &\\ 
\hline 
\end{tabular*}
\end{table}
Table~\ref{tab:ex} compares the schematic orbits of Fig.~\ref{fig:su3}
with the ones obtained by diagonalizing $2q^{20}$ associated to
``true'' $2q\cdot 2q$ and not one of its variants. The two bottom
lines give the cumulated values after filling up to i-th orbit with 2
neutrons and 2 protons.  Thus for 12 particles in $r_3$ and 4 in $gds$
we find \q=30.20+27.32=50.52. This table is the relevant one for
prolate states.

Quasi SU3 strongly prefers prolate solutions as Fig.~\ref{fig:su3}
makes clear: it is more advantageous to fill orbits from below than
from above.  

\begin{table}[t]
\caption{\label{tab:gr3}
 {\sf Top}: Intrinsic prolate and oblate
  quadrupole moments \q for $\nu$ particles in the 0$g_{9/2}$
  orbit (N=Z). 
   {\sf Bottom}: pseudo-SU3 \q for $\mu$ prolate particles
  ({\sf p-p}) or $\mu$ prolate holes ({\sf p-h}), -\q for $\mu$ oblate particles
  ({\sf -(o-p)}) or  $\mu$ oblate holes ({\sf -(o-h)})
}
\begin{tabular*}{\linewidth}{@{\extracolsep{\fill}}|c|cccccccc|}
\hline
 $\nu$& 2 &4  &6 &8& 10&12&14&16\\ 
\hline
prol & 5.33& 10.66& 14.66& 18.66& 20& 21.33 &  18.66& 16 \\
-obl & 8& 16& 18.66& 21.33& 20& 18.66& 14.66&10.66\\ 
\hline
\hline
 $\mu$  & 2 &4  &6 &8& 10&12&14&16\\
\hline 
{\sf p-p}; {\sf -(o-h)} &10.12&20.24&23.04&25.88&28.05&30.20&25.46&20.72\\
\hline
{\sf p-h}; {\sf -(o-p)} &5.22&10.44&15.66&20.72&25.46&30.20&28.04&25.88\\
\hline
\end{tabular*}
\end{table}

 \subsubsection{Single orbit quadrupole\label{sec:single}}
 When the $g$ orbit becomes sufficiently depressed with respect
to its $ds$ partners their influence can be neglected and we move
to the single $j$ orbit regime with quadrupole moments given by
\begin{equation}\label{eq:singlej}
Q_0=2\langle r^2C_2\rangle =\sum_m(p+3/2)\frac{j(j+1)-3m^2}{2j(j+1)}
\end{equation}
which shows that, before midshell, filling large $m$ values (negative
$Q_0$) is favored. The situation is reversed after midshell. Though
the notion of shape is questionable in this case, states with positive
and negative $Q_0$ will be referred to as prolate and oblate
respectively.

Table~\ref{tab:gr3} collects the possible values of \q for the
$g_{9/2}$ orbit and the $r_3$ space where one may wish to speak in
terms of holes rather than particles, and the table allows for all
possibilities. For example, under  $\mu$=8 we find that \q=25.88 
for prolate particles, 20.72 for prolate holes, -25.88 for oblate
holes and -20.72 for oblate particles.

 To guarantee a {\em bona fide} intrinsic state, $Q_0$ must coincide
 with the values extracted either from the spectroscopic quadrupole
 moment ($Q_{0s}$) 
\begin{gather}
Q_{spec}(J)=<JJ\vert3z^2-r^2\vert
JJ> \nonumber \\ 
Q_{0s}=\frac{(J+1)\,(2J+3)}{3K^2-J(J+1)}\,Q_{spec}(J), \quad K\ne1  
\label{bmq}
\end{gather}
 for Bohr Mottelson rotors, or the corresponding B(E2) transitions ($Q_{0t}$)
%%%%\vspace*{-.4cm}
\begin{gather}
 B(E2,J\;\rightarrow\;J-2)=\nonumber \\
 \frac{5}{16\pi}\,e^2|\langle 
JK20| J-2,K\rangle |^2 \, Q_{0t}^2\quad K\ne 1/2,\, 1
\label{bme2} 
\end{gather}
 The condition $Q_0\approx Q_{0s}\approx Q_{0t}$ is well fulfilled by
 SU3 states and its variants. ($Q_{0s}$ may be tricky though, as
   it is more sensitive to details than $Q_{0t}$. For an example refer
   to section~\ref{sec:zn60} ).

\section{Computational Strategy. SU3-Nilsson selfconsistency}\label{sec:comp} 
The guiding idea is that once quadrupole dominance sets in, the
wavefunctions are basically given by the quadrupole force which is
quite immune to single particle details. In other words \q varies
little. Our aim is to  estimate \q and understand the
reason for its stability.

We shall be interested in even $N=Z=$ 28 to 48 nuclei. Full $pf$
diagonalizations are possible but their interest is restricted to the
lightest species. For $r_3g$ exact calculations are also possible that
account for oblate states. The JUN45 interaction~\cite{jun45} will be
used throughout the region. Though the $r_3g$ space is of limited
relevance, the exact calculations  will serve as a test of our simple
models. For the more collective prolate states the full $r_3gds$ space
is necessary and exact calculations are not presently feasible, so we
shall introduce a selfconsistent version of Nilsson's model that
reduces to quasi and pseudo SU3 in the absence of a central
field~\cite{gz}.

\subsection{Example of naive $\bm{BE2}$ estimate}\label{sec:ex}

For SU3 the correct value of $Q_0$ to be used in Eqs.~(\ref{bmq},
\ref{bme2}) is $Q_0=$(\q+3)$b^2$~\cite{su3,scissors} with \q given in
Tables~\ref{tab:ex} or~\ref{tab:gr3}. In what follows we adopt this
form in all cases. 

The procedure is simple: use the tables to match oblate pseudo SU3
states in $r_3$ to oblate states in $g$ and prolate pseudo SU3 states
in $r_3$ to prolate quasi SU3 states in $gds$.  For instance: choose
16 particles and decide that we are interested in $^{72}$Kr
configurations with 12 particles in $r_3$ and 4 above. From the tables
we have for \q the following possibilities:

%\noindent 
\begin{quote}

Oblate
\vspace*{-.15cm}

\q= --30.2 for m=12 in pseudo, 
\vspace*{-.1cm}

\q= --16 for  n=4 in $g$. 
\vspace*{-.1cm}

Total $Q_0/b^2$= -(30.2+3+16)= -49.2

Prolate 
\vspace*{-.15cm}

\q= 30.2 for m=12 in pseudo, 
\vspace*{-.1cm}

\q=27.32  for n=4 in quasi. 
\vspace*{-.1cm}

Total  $Q_0/b^2=$ 30.2+3+27.32=60.52

Recover dimensions through 
\vspace*{-.1cm}

$b^2\approx 41.4/$\hw fm$^2$,~~ \hw=$45A^{-1/3}-25A^{-2/3}$
\end{quote}

\noindent
Now assume a conventional 2\hw scalar effective charge,
\mbox{$e_0=e_{\nu}+e_{\pi}=2$} {\em chosen throughout in what
  follows}. Then, for $A=72$, $b^2=4.42$ \ef, we have
$Q_0\approx -$217 \ef (oblate); 267 \ef (prolate).

The 2\hw effective charge is caused by coupling states in a major HO
shell to the giant quadrupole resonance. A rigorous derivation leads
$e_0$=1.77~\cite{mdz}, a number to be preferred~\cite{crawford_eff},
and shown in parenthesis below. Using \mbox{\bet$=[Q_0]^2/50.3$} from
Eq.~(8) leads to

\noindent
\mbox{\bet$\approx$ 936 (725) e$^2$fm$^4$} for oblate; 

\noindent
\bet$\approx$1422 (1101)  e$^2$fm$^4$ for prolate.

When working in EI or EEI spaces it becomes necessary to account for
0\hw polarization effects. In our case due to coupling
to the lowest $J=2^+$ state in $^{56}$Ni. The effect will be estimated
later leading to $e_0\gtrapprox 2$.

\subsection{\tcn{Nilsson revisited: the MZ equations}\label{sec:MZ}}

\tcn{The estimates above neglect single particle effects. To account for
them demands solving the Schr\"odinger equation for the quadrupole
force in the presence of a central field, a task as hard as the
general problem. In reference~\cite{a4749} Mart\'{\i}nez Pinedo and
Zuker (MZ) proposed to reduce it, by linearization, to a Nilsson type
Hamiltonian. That this should be possible seems obvious but the
implementation is not trivial. Because of a subtlety that was missed at
the time, the project was left unfinished. We retake it.}

We would like to solve
\begin{gather}
  \label{hmq}
  H_{mq}= \sum \epsilon_in_i
 -\hbar\omega\kappa\left(\frac{2q_p}{{\cal N}_{2q,p}}+
\frac{2q_{p+1}}{{\cal N}_{2q,p+1}}\right)^2\\ 
 \label{N2q}
{\cal N}_{2q,p}^2=\sum (2q_{rs})^2=\frac{5}{2}\sum_{k=0}^p (k+1)(2p-3k)^2
\end{gather}
where we have borrowed from~\cite{mdz} the normalized form of the
quadrupole force that emerges naturally when it is extracted from a
realistic interaction ($q_p$ is the quadrupole operator in major shell
$p$, the square stands for scalar product). This form ensures that
$\kappa\approx 0.22-0.25$ is a universal constant that demands a 30\%
renormalization due to coupling to the 2\hw quadrupole degrees of
freedom~\cite{mdz}. \tcn{It also ensures that nuclei do not become
  needles, thus solving the crippling problem of the naive quadrupole
  force~\cite{bk68}}. In all that follows we have fixed $\kappa=0.3$.

 To prepare for linearization replace $q$ by $q_{20}$ operators
 \tcn{(notice that we use sometimes $q_{20}$ instead of $q^{20}$ for
   typographical reasons)} .
\begin{gather}
  \label{hmq0}
  H_{mq0}= \sum \varepsilon_in_i
 -\hbar\omega\kappa\left(\frac{2q_{20,p}}{{\cal N}_{2q_{20,p}}}+
\frac{2q_{20,p+1}}{{\cal N}_{2q_{20,p+1}}}\right)^2\\ 
{\cal N}_{2q_{20,p}}^2=\sum (2q_{20,rs})^2=\sum_{k=0}^p (k+1)(2p-3k)^2
 \label{N2q20}
\end{gather}
Note that Eq.~(\ref{N2q20}) is obtained by summing the squares of the
levels in Fig.~\ref{fig:SU3}.

Now concentrate on a single space. The operation amounts to
replacing $2q\cdot 2q$ by $2q_{20}2q_{20}$, and demands some care
because $q_{20}$ is a sum of neutron and proton contributions
$q_{20}=q_{20}^{\nu}+q_{20}^{\pi}$. As calculations will be done for
each fluid separately, the correct linearization for the neutron
operators, say, is:

\vspace{6pt}

 $q_{20}\, q_{20}\to
q_{20}^{\nu}\langle q_{20}^{\nu}+2q_{20}^{\pi}\rangle \approx
3q_{20}^{\nu}\langle q_{20}^{\nu}\rangle$ if
$\langle q_{20}^{\nu}\rangle\approx \langle q_{20}^{\pi}\rangle$

\vspace{6pt}

\noindent
\tcn{leading to the Mart\'{i}nez Zuker (MZ) equation}

\begin{gather}
 H_{mq0}=\sum \varepsilon_in_i -\frac{3\hbar \omega \kappa}{{\cal
      N}^2_{2q_{20},p}}\langle 2q_{20}^{\nu}\rangle 2q_{20}^{\nu}
\label{hmql1}
\end{gather}
The subtlety missed in~\cite{a4749} was the need to change ${\cal
  N}_{2q,p}$ into ${\cal N}_{2q_{20},p}$ in going from Eq.~(\ref{hmq})
to Eq.~(\ref{hmq0}), thus making it impossible to discover the proper
way to proceed which now can be implemented~\cite{gz}.

To find the proper generalization of Eq.~(\ref{hmql1}) note that in
the full space $q$ becomes a sum of four contributions
$q_{20}=q_{20}^{\nu u}+q_{20}^{\pi u}+q_{20}^{\nu d}+q_{20}^{\pi d}$
($u=gds$, $d=r_3$).  By repeating the arguments leading to
Eq.~(\ref{hmql1}) and setting ${\cal N}_i={\cal N}_{2q_{20},i}$
\tcn{leads to the general MZ equation}

\begin{gather}\nonumber
H_{sp}-4\hbar \omega \kappa \frac{q_4^{\nu}}{{\cal N}^2_4}\left(\langle
q_4^{\nu}\rangle +\langle2q_4^{\pi} \rangle+\langle
2q_3^{\nu}\rangle\frac{{\cal N}_4}{{\cal N}_3}+\langle
2q_3^{\pi}\rangle\frac{{\cal N}_4}{{\cal N}_3} \right)\\ \approx
H_{sp}-4\hbar \omega \kappa \frac{q_4^{\nu}}{{\cal N}^2_4}\left(
3\langle q_4^{\nu} \rangle+6\langle q_3^{\nu}\rangle \right)\nonumber\\= H_{sp}-\beta\hbar
\omega \kappa \frac{\langle 2q_4^{\nu}\rangle}{{\cal N}^2_4}2q_4^{\nu} 
\label{hmql2}
\end{gather}
where we have introduced a boost factor $\beta$, set
$q_i^{\nu}=q_i^{\pi}$ and used the correct numbers from
Eq.~(\ref{N2q20}) ${\cal N}_{2q,3}=\sqrt{90\times 2.5}\approx15$,
${\cal N}_{2q,4}=\sqrt{210\times 2.5}\approx 23$.
to approximate ${\cal N}_4/{\cal N}_3=22.91/15\approx
1.5,$ 

As the \q ranges will be $\langle 2q_3 \rangle\approx 30$ and $\langle
2q_4 \rangle=27-55$, the modest value of $\beta=3$ in
Eq.~(\ref{hmql1}) will increase to about $\beta=9$-12, but the work
involved in solving Eqs.~(\ref{hmql1}) and (\ref{hmql2}) is identical.

Let us examine the steps involved.

\begin{figure}[h]
\begin{center}
  \includegraphics[width=1.0\columnwidth]{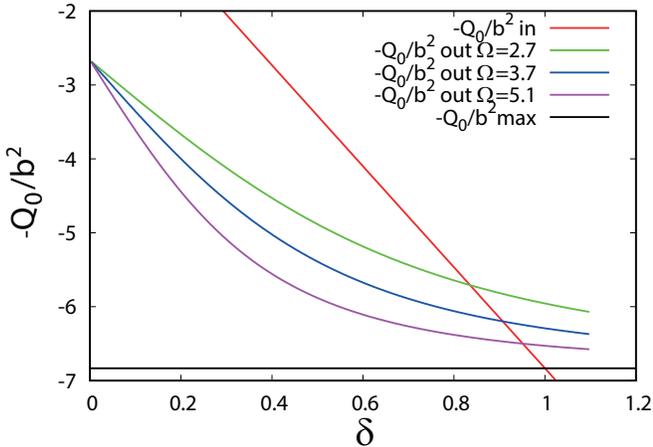}
\end{center}
\caption{\label{fig:q0_self} (color online) Calculation of $\langle
  2q_{20}\rangle$ in $(gds)^4$. Input values in red. Output values
  obtained solving Eq.~(\ref{hmql1}) for $\varepsilon_i=0$,1,2 MeV for
  $i=g,d,s$ respectively. When input and output values coincide,
  selfconsistency is achieved \ie when lines cross, which happens at
  abscissae 0.83, 0.91, 0.95 for $ \Omega=\hbar \omega \kappa=$ 2.7,
  3.7, 5.1 respectively.  At $\delta=0$ \q=2.666 corresponds to one
  prolate $g$ orbit (from Table~\ref{tab:gr3})}
\end{figure}
a) Eq.~(\ref{hmql1}) is solved setting as inputs $\langle
2q_{20in}\rangle=\delta \langle 2q_{20max}\rangle$, which for $\delta=1$
yields the maximum value of $\langle 2q_{20}\rangle$ (the one obtained
at $\varepsilon_i=0$). The resulting eigenvalue can be written as
\begin{gather}
 \label{hmqlout}
  E(\delta)=\langle H_{sp}\rangle -\frac{3\hbar \omega
    \kappa}{{\cal N}^2_{2q_{20}}} \delta  \langle  2q_{20max}\rangle \langle
  2q_{20out}\rangle
\end{gather}

b) Extract $\langle 2q_{20out}\rangle$, use it as next input and
iterate until $\langle 2q_{20in}\rangle=\langle 2q_{20out}\rangle$.
Fig.~\ref{fig:q0_self} sums up the procedure.

c) Guess energies. The comparison of the resulting $\langle
2q_{20out}\rangle=Q_0/b^2$ with exact results turns out to be
systematically satisfactory. Some examples are given in
Section~\ref{sec:ener}.  As a reasonable estimate of $Q_0/b^2$ amounts
to a good guess of intrinsic state from which the energy could be
extracted by taking the expectation value of $H_{mq}$ in
Eq.~(\ref{hmq}) but it is more instructive and simpler to stand by our
basic assumption that $2q_{20}$ is an acceptable quantum number and
rely on the exact SU3 result as a guide.
\begin{gather}
E=-\frac{\hbar \omega \kappa}{{\cal N}_{2q}^2}\left(2\lambda(2\lambda+6)
+2\mu(2\mu+6)+4\lambda\mu-3L(L+1)\right)\nonumber
\end{gather}
where $\lambda$ and $\mu$ are the difference in quanta in the z and x,
and x and y directions respectively. This result is valid for the
$q\cdot q$ force that contains one and two body parts. As we are not
interested in the former we expect modifications if they are
neglected.  Moreover, we shall restrict the energy estimates to
$(\lambda 0)$ representations because the only obviously correct
identification in the absence of external monopole fields is
2$\lambda=$\q. The idea is to assume that the quadrupole contribution
to the energy keeps this form using the calculated \q value.

The proposed estimates are as follows
\begin{gather}
   BE2=B(E2:2^+\to 0^+)=((\langle 2q_{20}\rangle +3)b^2)^2/50.3\nonumber\\
   Q_{s}=Q_{(J=2)s}=(\langle 2q_{20}\rangle +3)b^2/3.5\nonumber\\
  E= \langle H_{sp}\rangle
  -\frac{\hbar \omega \kappa}{{\cal N}_{2q}^2}\langle
  2q_{20}\rangle(\langle 2q_{20}\rangle +\zeta)
  \label{hmqen}
\end{gather}
For $Q_s$ and \bet we use Eqs.~(\ref{bmq}-\ref{bme2}). The norms are
those of the full quadrupole interaction \ie ${\cal
  N}_{2q}=\sqrt{2.5}{\cal N}_{2q20}$. The parameter $\zeta$ in the
form of $E$ should be 6 if the SU3 analogy held perfectly. However, as
hinted above and made evident in next section~\ref{sec:ener} this is
not possible and $\zeta$ must be viewed as an artifact to estimate
uncertainties in the guessed energies.

Finally, let us propose a generalization of Eq~(\ref{hmqen}) for the the
energy of a $r_3^k(gds)^l$ configuration, and write explicitly $Q_s$
and \bet  Calling $\langle
2q_{20(i)}\rangle=Q_i$, we have
 \begin{gather}
   BE2=B(E2:2^+\to 0^+)=((Q_3+Q_4 +3)b^2)^2/50.3\nonumber\\
   Q_{s}=Q_{(J=2)s}=(Q_3+Q_4+3)b^2/3.5\nonumber\\
  E= H_m   
-\hbar \omega \kappa(\frac{Q_3}{15}+\frac{Q_4}{23})
(\frac{Q_3+\zeta}{15}+\frac{Q_4+\zeta}{23})      
\label{hmq2en}
\end{gather}
where the monopole term $H_m$ subsumes the evolving behavior of the single particle  
fields to be discussed in Section~\ref{sec:central}.

\subsection{Pseudo and Quasi SU3 as exact symmetries}\label{sec:ener}
According to SU3, $^{28}$Si has a prolate-oblate degenerate ground
state corresponding to the $(\lambda, \, \mu)=$ (12,0) and (0,12)
representations. This holds for the full $q\cdot q$, \ie including
both its one and two body terms. If the former are ignored we obtain
the results in the right upper panel of Table~\ref{tab:si28se68},
which show no signs of an exact degeneracy. The estimated energy
using $\zeta=0$ in Eq.~(\ref{hmqen}) is about 5\% larger than the
exact one. Nearly perfect degeneracy is achieved with the monopole
free $q\cdot q$---\ie with all centroid averages set to 0---in the
left upper pannel and the estimated energy with $\zeta=3$ is now some
5\% too small.
\begin{table}[h]
\caption{\label{tab:si28se68}$q\cdot q$ calculations in $^{28}$Si($(sd)^{12}$)
  (upper panels) and $^{68}$Se($r^{12}_3$) (lower panels). To the left
$q\cdot q$ is made monopole free. ``Int'' stands for intrisic values
  defined in Eq.~(\ref{hmqen}), with $\zeta=3$ in the left panel and 0
in the right one. Absolute energies given for the ground state,
excitation energies for the other states (MeV).}
\begin{tabular*}{\linewidth}{@{\extracolsep{\fill}}|cccc|cccc|}
\hline
J & $E$ &$Q_{s} $&$BE2$ &J & $E$ &$Q_{s} $&$BE2$ \\
\hline
  0&  -27.26329&          &          &     0& -22.044&	       &\\		   
  0&    0.00192&          &          &     2&   0.958&  -26.330& 166.979\\
  2&    0.91714&   -0.7785&   167.299&     0&   1.646&	       & \\	   
  2&    0.91730&    0.7785&   167.301&     2&   2.494&   26.323& 166.903\\
       \hline
Int&  -26.20   & -26.40   &   169.74&Int&  -23.23   & -26.40   & 169.74\\
\hline
\hline
   0&  -14.97822&           &         &  0& -12.176&	      &	\\	
   0&    0.00042& 	    &	      &  2&   0.533& -42.087&  426.802\\
   2&    0.50677&  -2.8707  &  427.591&  0&   0.996& 	      &	\\	
   2&    0.50683&   2.8708  &  427.590&  2&   1.467&  42.117&  426.915\\
\hline
Int&   -13.99& 41.21& 413.64&Int&   -12.73& 41.21& 413.64\\
\hline
 \end{tabular*}
\end{table}
The story repeats itself in the lower panels for $^{68}$Se: a
remarkable result establishing that pseudo-SU3 behaves as an exact
symmetry in this case. \tcn{A puzzling result since we are using the true
$q\cdot q$ potential whose matrix elements coincide in magnitude with
their pseudo counterparts but have different sign structure. So much
so that their overlaps (in the sense of~\cite[Eq.(44)]{rmp}) nearly
vanish.}

Let us draw some conclusions.
\begin{itemize}
\item  Energies are very sensitive to monopole behavior but \bet rates
  are not.
\item When bands---with equal \bet and opposite $Q_s$---cross, they
  mix leading to unchanged \bet and cancellation of $Q_s$. Note that
  this could happen through small ``impurities'' in the
  Hamiltonian. If the symmetry were exact, the Lanczos algorithm used
  in the diagonalizations could not break the degeneracy.
\item Pseudo SU3 appears to be close to an exact symmetry.
\end{itemize}
\subsection{Checks}\label{sec:checks}
Allowing the single particle energies to vary produces more stringent
tests of the estimates in Eq.~(\ref{hmqen}). Numerous calculations
done for the $r_3^n$ and$(gds)^n$ spaces lead to results that
are well summarized by the examples in Table~\ref{tab:gds8-12}.
\begin{table}[h]
\caption{\label{tab:gds8-12}Monopole-free $q\cdot q$ calculations in
  $(gds)^{8-12}$. Single particle energies in MeV:
  $\varepsilon_i$=[0.0, 0.0 and 0.0](e0) and [0.0, 1.0, and 2.0](e1)
  for $i=g,\, d,\, s$ respectively. ``Int'' stands for intrisic values
  defined in Eq.~(\ref{hmqen}) with $\zeta=3$.}
\begin{tabular*}{\linewidth}{@{\extracolsep{\fill}}|cccc|cccc|}
\hline
J & $E$ &$Q_{s} $&$BE2$ &J & $E$ &$Q_{s} $&$BE2$ \\
\hline
   \multicolumn{4}{|c|}{\tcn{$(gds)^8$} e0}&\multicolumn{4}{c|}{$(gds)^8$ e1}\\
\hline   
   0& -12.977&        &	       &	0&  -8.976&	   &	    \\
   2&   0.113& -59.735& 857.959&        2&   0.103& -57.313& 795.566\\
\hline
Int & -12.34 & -59.98 & 876.05 &       Int&  -8.28&  -57.50&  805.07 \\
\hline 
   \multicolumn{4}{|c|}{$(gds)^{12}$ e0}&\multicolumn{4}{c|}{$(gds)^{12}$ e1}\\
\hline   
  0&   -17.894&		&        &     0&  -12.641&	    &  \\		 
  2&     0.125& -65.296& 1161.574&     2&    0.136& -65.609&  1065.721\\
\hline
Int&   -15.84 &  -69.16& 1165.04&     Int&    -8.63  &  -67.49&   1109.34\\ 
\hline
    \end{tabular*}
\end{table}

 In all cases the selfconsistent calculations do well for the
 quadrupole properties. The estimated energies in $(gds)^8$ are also
 satisfactory, but they fall short for $(gds)^{12}$ which is triaxial
 because the platform at \q=-1.5 in Fig.~\ref{fig:su3} is not fully
 filled, leading to $\mu\ne 0$ while $E$ in Eq.~(\ref{hmqen}) is
 designed for $\mu=0$ only.

\section{N=Z nuclei} 
Granted the benefit of some hindsight, a reading of Fig.~\ref{fig:su3}
suggests three regimes for $N=Z$ nuclei from $^{56}$Ni up to
$^{96}$Cd. Note that placing the $gds$ ``quasi'' orbits  on top of the
``pseudo'' $r_3$ ones was designed to facilitate such a reading.

\vspace{0.15cm}
\noindent
i) The $r_3$ pseudo SU3 nuclei. They fill orderly the three lowest
levels in Fig.~\ref{fig:su3}: $^{60}$Zn (analog of $^{20}$Ne in the
$sd$ shell, a mild rotor), $^{64}$Ge (analog of $^{24}$Mg, a rotor
exhibiting a $\gamma$ band, as expected whenever orbits are not all
filled at a given level), $^{68}$Se (analog of $^{28}$Si, with
degenerate prolate and oblate bands). While SU3 dominance is largely
frustrated in the $sd$ shell, here it is expected to hold well because
of the near degeneracy of the single particle orbits.  This region
makes it possible to study the full $pf$ to $r_3$ reduction: A unique
opportunity to validate the notion of model space and in particular
the assumption that $^{56}$Ni can be treated as a closed shell. As for
$r_3g$ calculations~\cite{kaneko}, they add little to the $r_3$ ones.

\vspace*{0.15cm}
\noindent
ii) Coexistence from $^{72}$Kr to $^{84}$Mo.  For 12 particles, \ie
$^{68}$Se, \q reaches a maximun in $r_3$ (see last lines of
tables~\ref{tab:ex} and ~\ref{tab:gr3}). Adding particles to the
pseudo orbits leads to a loss while adding them to the quasi orbits
leads to a gain. By filling the quasi orbits, well deformed prolate
states can be constructed for 4, 8, 12 and 16 particles whose
quadrupole energy will overcompensate the monopole (\ie single
particle) losses. Oblate states very close in energy can also be
found, leading to coexisting bands. The prolate and oblate states
demand $r_3gds$ and $r_3g$ spaces respectively. The associated
dimensionalities exceed $10^{14}$ for the former and $10^{10}$ for the
latter---still large but feasible. Therefore we shall rely on a
generalization of the simplified approach of Section~\ref{sec:ener}
for both deformations and check the oblate results via exact
diagonalizations.  For studies of the region see~\cite{vampir,64-84}.

\vspace*{0.15cm}
\noindent
iii) The $r_3g$ nuclei $^{88}$Ru  $^{92}$Pd and  $^{96}$Cd. The second has been
measured recently~\cite{ceder:10} and postulated as candidate for a new
form of boson aligned collectivity. We shall examine the
claim. The---still unknown---spectrum of $^{96}$Cd will be shown to be
probably closer to $sdg$ than to $r_3g$. 

\subsection{The $\bm {pf}$ to $\bm{r_3}$ reduction}\label{sec:red}
 
 Doubts may be raised about the doubly magic nature of $^{56}$Ni as its
first 2$^+$ is rather low and, depending on the effective interaction
used ({\sc kb3g, gxpf1a})~\cite{kb3g,gxpf1a} the closed shell component
amounts to only 60-70\%. However, it is in the nature of the shell
model to recognize that there may be a difference between the
potentially complicated structure of a state and its simple behavior.
As a first hint of what is expected of magic nuclei we refer to 
Figures 1--5 in~\cite{alpha}: at magic numbers, two-neutron and two-proton
separation energies exhibit {\em systematic} jumps. Clearly the case
for $N$ or $Z=\,$28, and {\em a fortiori} for $^{56}$Ni. Not for
occasional candidates such as $N=56$ which is magic only for $Z=40$.
  
For our present purpose the state of interest is the head of the 4p-4h
rotational band. According to Eq.~(\ref{eq:singlej}) four holes in the
0f$_{7/2}$ orbit give a prolate contribution of 12b$^2$ to the
intrinsic quadrupole moment while four pseudo SU3 particles in $r_3$
contribute with $\approx$ 22b$^2$, adding up to 34b$^2$ in
agreement with 32b$^2$ from a full 4p-4h $pf$-shell calculation. A
first example of the use of our schematic \tcn{coupling schemes.}

\vspace{-.5 cm}

\subsubsection{0\hw polarization}\label{sec:0hw}

\vspace{-.2 cm}
The most important characteristic of a doubly magic nucleus is that it
defines a before and an after. Before $^{56}$Ni, nuclei are basically
of $f$ type. Beyond, they are at first of $r_3$ type until  the
extension to $r_3gds$ spaces becomes imperative. To treat $^{56}$Ni as
a core, the Hamiltonian and transition operators have to be
renormalized. The dominant mechanism involves coupling to the low
lying $2^+$ state, leading to three-body forces and two-body effective
transition operators~\cite{Pb} (\ie state dependent effective charges)
whose neglect, as stressed in ref.~\cite{mdz}, is ``common but bad
practice''. Short of a rigorous treatment we chose the following
expediencies:
\begin{itemize}
\item For the energies we assume that {\sc jun45}~\cite{jun45}
  provides a reasonable approximation to the effective Hamiltonian. To
  fix ideas: in~\cite{mdz} it is shown that for the quadruplole
  component of the bare realistic forces the 2\hw effects demand a
  30\% boost (consistent with what is known about phenomenological
  interactions). As a consequence the effective $q\cdot q$ amounts to
  about 50\% of the total interaction. In the case of {\sc jun45} it
  jumps to over 75\%, indicating a strong contribution due to 0\hw
  mechanisms.
\item For the transition operators we proceed by brute force,
  estimating effective charges by comparing full $pf$ transitions
  rates to those obtained in the $r_3$ or $r_3g$ spaces.
\end{itemize}

%\vspace*{-1.cm}

\subsubsection{$^{60}$Zn, more on magicity } \label{sec:zn60}
To check that $^{60}$Zn is properly described by $r_3^4$
configurations we do a full $pf$ diagonalizations which involves
2.292.604.744 M=0 Slater determinants. The story is told in
Table~\ref{tab:zn60}. A calculation in the $r_3$ space, using a pure
quadrupole-quadrupole interaction gives values in the range 24b$^2$.
As expected we have good rotational features including $J(J+1)$
spacings. The full $pf$-shell calculation using the {\sc kb3gr}
interaction~\cite{kb3gr} accounts well for the experimental
spectrum. The $J(J+1)$ spacings are gone but this is of little
consequence. As abundantly emphasized in ~\cite{zrpc} what matters is
the wavefunction \ie the quadrupole moments. The spectrum may be
sensitive to details detected in first order perturbation theory that
do not change the structure of the state. And the message from
Table~\ref{tab:zn60} is that the quadrupole moments of the huge
calculation and the modest one are compatible, to within a crucial
caveat: The full $pf$ space leads to $Q_{0t}$ values that are about
1.36 times bigger than the $r_3$ ones.  As the coupling is mediated
basically by the $p_{3/2}f_{7/2}^{-1}$ jumps the renormalization
decreases as the $p_{3/2}$ orbit gets filled thus blocking the jumps.
The results hardly change when {\sc jun45} is used instead of $q\cdot
q$ in the $Q_{0t,qq}$ column of Table~\ref{tab:zn60}, 23 goes to 20.8,
increasing the enhancement factor $F$ from 1.36 to 1.48. The
calculated spectrum---though still dilated---comes closer to the
experimental one.

Note that the evolution of $Q_{0s}$ and $Q_{0t}$ are quite different.
In general the two quantities will be approximately equal only in the
case of well developed rotors. More often than not $Q_s$ is very
sensitive to details, while $Q_t$ is close to the predictions from
Tables~\ref{tab:ex} and \ref{tab:gr3}.
 \begin{table}[h]
\caption{\label{tab:zn60} Properties of the yrast band of $^{60}$Zn
  ($E$ in MeV, $Q$ in units of b$^2$).Calculations: full $pf$ with
  {\sc kb3gr}; \tcn{and} $r_3 $ with $q\cdot q$.}
\begin{tabular*}{\linewidth}{@{\extracolsep{\fill}}|cccccccc|}
\hline
J & $E_{exp}$ &$ E_{qq}$&$ E_{pf}$ &$Q_{0s,qq}$&$Q_{0s,pf}$&$Q_{0t,qq}$&$Q_{0t.pf}$\\
\hline  
 2$^+$  & 1.00 & 1.00 & 1.07 & 24 & 22  &  23 &   31   \\ 
 4$^+$  & 2.19 & 3.34 & 2.31 & 23 & 25  &  22 &   30   \\ 
 6$^+$  & 3.81 & 7.03 & 4.06 & 23 & 14  &  19 &   31   \\ 
       \hline
    \end{tabular*}
\end{table}
	
It is worth mentioning that $^{60}$Zn has a superdeformed excited band
at relatively low energy with $Q_0$=~67(6)~b$^2$ \cite{sd60zn}. From
Tables~II and III two prolate candidates emerge with configurations
$f^{12}r_3^4(gds)^4$ and $f^{12}(gds)^8$. Both are consistent with
observation.
\subsubsection{$^{64}$Ge}\label{sec:ge64}
\begin{table}[h]
\caption{\label{tab:ge64} Properties of low lying states in $^{64}$Ge,
  Energies in MeV, $B(E2)$ in \eff. Calculations: full $pf$ with {\sc
    gxpf1a}~\cite{ge64}; $r_3g$ with {\sc jun45} and $r_3 $ with
  $q\cdot q$.}
\begin{tabular*}{\linewidth}{@{\extracolsep{\fill}}|c|ccccc|}
\hline
$J^\pi$ & & Exp&$pf$&$r_3g$ &$q\cdot q$\\
\hline
2$^+_1$& $E_x$ & 0.90 & 0.94 & 0.86 & 0.50\\
2$^+_1$& $Q_s$ & & -18.6 & -24.4 & 5.03 \\
& $B(E2:2_1^+\rightarrow 0_1^+)$ & 410(60)& 406 & 251 & 300 \\
2$^+_2$& $E_x$ & 1.579 & 1.56 & 1.27 & 0.55 \\
2$^+_2$& $Q_s$ & & 18.5 & 23.3 & -5.42 \\
& $B(E2:2_2^+\rightarrow 2_1^+)$ & 620(210)& 610 &182 & 479 \\
& $B(E2:2_2^+\rightarrow 0_1^+)$ & 1.5(5)& 14 & 13 & 39 \\
4$^+_1$& $E_x$ & 2.053 & 2.00 & 2.16 & 1.61  \\
& $B(E2:4_1^+\rightarrow 2_1^+)$ & & 674 & 314 & 390 \\
\hline
\end{tabular*}
\end{table}
For $^{64}$Ge the diagonalization of the $q\cdot q$ interaction in the
$(r_3)^8$ space yields the expected results for an (84) SU3
representation with nearly degenerate 2$^+$ states---with $Q_0$ of
equal magnitude and opposite signs---\tcn{corresponding} to the $K$=0 and 2
ground state and $\gamma$ bands respectively, and \bet of about
300 \eff.  Table~\ref{tab:ge64} proposes a comparison of $q\cdot q$ and
{\sc jun45} results---in $r_3$ and $r_3g$ spaces respectively---with data,
well reproduced by {\sc gxpf1a} calculations~\cite{ge64}. Using as reference
the \bet values it is found that in going from $r_3$ to $pf$ the
enhancement factors $F$ are 1.62 (for {\sc jun45}) and 1.23 (for $q\cdot q$).
\subsubsection{$^{68}$Se: The double platform}\label{sec:se68}
The structure of $N=Z$ even nuclei from $A=$72 to 84 will be described
by piling up $(gds)^4$ blocks on top $r_3^{12}$, \ie on top of either
the oblate and prolate ground state bands---corresponding to the (12 0)
and (0 12) SU3 representations---of $^{68}$Se which becomes a
common ``double platform'' (refer to Fig.~\ref{fig:su3}).  Hence the
importance of this nucleus to fix the $e_0$ effective charge.

%\begin{table}[h]
%\caption{\label{tab:se68} Properties of low lying states in $^{68}$Se,
%  Energies in MeV, $B(E2)$ in \eff. Calculations: full $pf$ with {\sc
%    gxpf1a}~\cite{ge64}; $r_3$ with {\sc jun45} \tcn{and}  $q\cdot q$\tcn{;} and full
%  $pf$ with {\sc kb3gr} ($PF$). The experimental $0^+_2$ energy is a
%  guess. Note inversion of 1 and 2 indexes for $q\cdot q$ which has
%  prolate ground state.}
%\begin{tabular*}{\linewidth}{@{\extracolsep{\fill}}|c|cccccc|}
%\hline
%$J^\pi$ & & Exp&$pf$&$r_3$ &$q\cdot q$&$PF$\\
%\hline
%0$^+_2$& $E_x$ & (1.19) & 0.72 & 0.96 & 0.78&1.44 \\
%2$^+_1$& $E_x$ & 0.85 & 0.71 & 0.54 & 0.53&1.10 \\
%2$^+_1$& $Q_s$ & & 23.9 & 37.7 & 41.7&2.92 \\
%& $B(E2:2_1^+\rightarrow 0_1^+)$ & 440(60)& 492 & 357 & 420&434 \\
%2$^+_2$& $E_x$ & 1.59 & 1.05 & 1.39 & 1.53&1.66 \\
%2$^+_2$& $Q_s$ & & -20.1 & -35.5 & -41.7 &-3.34\tcn{?}\\
%& $B(E2:2_2^+\rightarrow 2_1^+)$ & & 553 & 8 & 10&598 \\
%& $B(E2:2_2^+\rightarrow 0_2^+)$ & & 486 & 304 & 420&450 \\
%& $B(E2:2_2^+\rightarrow 0_1^+)$ & & 4 & 0.8 & 0.05&5.46\\
%4$^+_1$& $E_x$ & 1.94 & 1.66 & 1.61 & 1.77&2.10 \\
%4$^+_1$& $Q_s$ & & 59.8 & 46.6 & 53.5&64.13\\
%& $B(E2:4_1^+\rightarrow 2_1^+)$ & & 670 & 486 & 565&648 \\
%4$^+_2$& $E_x$ & 2.55 & 2.06 & 2.28 & 2.37&2.67\\
%4$^+_2$& $Q_s$ & & -52.5 & -45.7 & -53.6&-32.86\tcn{?} \\
%& $B(E2:4_2^+\rightarrow 2_2^+)$ & & 559 & 411 & 565&335\\
%\hline
%\end{tabular*}
%\end{table}
%

\begin{table}[h]
\tcn{
\caption{\label{tab:se68} Properties of low lying states in $^{68}$Se,
  Energies in MeV, $B(E2)$ in \eff. Calculations: full $pf$ with {\sc
    gxpf1a}~\cite{ge64}; $r_3$ with {\sc jun45}: $r_3$ with $q\cdot q$\tcn{;} and full
  $pf$ with {\sc kb3gr} ($PF$). The experimental $0^+_2$ energy is a
  guess.}
\begin{tabular*}{\linewidth}{@{\extracolsep{\fill}}|c|cccccc|}
\hline
$J^\pi$ & & Exp&$pf$&$r_3$ &$q\cdot q$&$PF$\\
\hline
0$^+_2$& $E_x$ & (1.19) & 0.69 & 0.96 & 0.79&1.42 \\
2$^+_1$& $E_x$ & 0.85 & 0.71 & 0.54 & 0.53& 0.96 \\
2$^+_1$& $Q_s$ & & 11 & 35 & -42 & 39 \\
& $B(E2:2_1^+\rightarrow 0_1^+)$ & 440(60)& 491 & 307 & 420 &409 \\
2$^+_2$& $E_x$ & 1.59 & 1.00 & 1.39 & 1.26&1.74 \\
2$^+_2$& $Q_s$ & & -8 & -33 & 42 & -16 \\
& $B(E2:2_2^+\rightarrow 2_1^+)$ & & 689 & 7 & 0.00 & 297 \\
& $B(E2:2_2^+\rightarrow 0_2^+)$ & & 499 & 262 & 420 & 223 \\
& $B(E2:2_2^+\rightarrow 0_1^+)$ & 0.3  & 4 & 0.7 & 0.00 & 10\\
4$^+_1$& $E_x$ & 1.94 & 1.66 & 1.61 & 1.77 & 1.86 \\
4$^+_1$& $Q_s$ & & 59 & 43 & -53& 63\\
& $B(E2:4_1^+\rightarrow 2_1^+)$ & & 590 & 419 & 565 & 810 \\
4$^+_2$& $E_x$ & 2.55 & 1.98 & 2.28 & 2.37&2.79\\
4$^+_2$& $Q_s$ & & -51 & -42 & 53&-14 \\
& $B(E2:4_2^+\rightarrow 2_2^+)$ & & 510 & 354 & 565& 154\\
\hline
\end{tabular*}}
\end{table}

From Table~\ref{tab:ex} the estimate $Q_0/b^2=\pm 33.2$ \ie \bet
$\approx 414$ \eff, consistent the $q\cdot q$ numbers in
Table~\ref{tab:se68}, which also collects {\sc jun45} results in
$r_3$, the full $pf$ {\sc gxpf1a} and {\sc kb3gr} ones (labeled $pf$
and $PF$ respectively) and data including the only experimentally
known \bet= 440(60)~e$^2$fm$^4$.

With the exception of the $B(E2:2_2^+\rightarrow 2_1^+)$ the
calculations in $r_3$ and $pf$ are quite consistent, with 
enhacement factors $F\approx$ 1.16 and 1.38 for the $q\cdot
q$ and {\sc jun45} numbers respectively. The {\sc kb3gr} interaction yields
somewhat better spectra than {\sc gxpf1a}, and similar quadrupole
properties except for the $J=2^+_2$ and $4^+_2$ states that are more
mixed for the latter. 

Using the 2\hw value $e_0=$1.77~\cite{mdz}, the 0\hw contribution
increases it to $e_0=1.77\sqrt{F}\approx 2.1\pm 0.1$. When $gds$
particles come into play their quadrupole operators will also couple
with the $J=2^+$ state in $^{56}$Ni, though more weakly due to larger
norm denominators (see Eqs.~(\ref{hmq} and~\ref{hmq2en})). It is
hoped that the associated suppression can be accommodated by the
proposed estimate.

\vspace{8pt}

The {\sc jun45} calculation \tcn{in $r3g$} leads to a ground state that is 60\%
0p-0h, 30\% 2p-2h and 10\% 4p-4h. As can be gathered from Tables I and
II these admixtures bring no extra oblate coherence but with the same
numbers, prolate contributions could make a difference in a full
$r_3gds$ calculation. Vampir calculations~\cite{vampirse} indicate
substantial oblate-prolate mixing in the ground state band. Further
data on this nucleus could be of interest.
\subsection{The central region: A=72 to 84}\label{sec:central} 
Let us recast $E$ in Eq.~(\ref{hmq2en}) so as to separate the two basic
contributions to the monopole term $H_m$. 
\begin{gather}
  E= \sum \varepsilon_{i=g,d,s}\langle n_i \rangle +  
l(\varepsilon_{g}-\varepsilon_{r_3})\nonumber\\
-\hbar \omega \kappa\left(\frac{Q_3}{15}+\frac{Q_4}{23}\right)
\left(\frac{Q_3+\zeta}{15}+\frac{Q_4+\zeta}{23}\right)  
\nonumber\\ 
 =hsp+l\epsilon_{gr}+E_q \label{hmq2enbis}
\end{gather}
where we have introduced the notations used in
Table~\ref{tab:central}---the core of this study--- which lists the
properties of the dominant and subdominant prolate and oblate states.
\begin{table}[h]
\caption{Properties of $r_3^k(gds)^l$ configurations. Total ($E$),
  quadrupole ($E_q$) and single particle ($hsp$) energies from
  Eq.~(\ref{hmq2enbis}) with $\zeta=0$, in MeV; quadrupole moment
  $Q_4=\langle 2q_{20(4)}\rangle$; $BE2=$\bet in \eff from
  Eq.~(\ref{hmq2en}); $\beta=8$ in Eq.~(\ref{hmql2}). For prolate
  states $Q_4$ is the calculated one. For oblate states the space is
  $r_3^k(g)^l$ so $\epsilon_{r_3}=0.0$ and $Q_4$ (not shown) is from
  Table~\ref{tab:gr3}. $Q_3$ is always from
  Table~\ref{tab:gr3}. Energies of triaxial states in boldface. Single
  particle energies in MeV: $\varepsilon_i$=0.0, 3.0 and 4.0 [0.0,
    4.0, and 5.0] for $i=g,\, d,\, s$ respectively, and
  $\epsilon_{gr}=2.5$ [2.0]. Numbers in square brackets apply to the
  last two lines only.
  \label{tab:central}}
\begin{tabular*}{\linewidth}{@{\extracolsep{\fill}}|ccccccccc|}
\hline
  {\tcn  k}  & l & {\tcn A} &  $E$    &   -$E_q$    &   $hsp$   &$Q_3$&    $Q_4$ &    $BE2$    \\ 
\hline   
   12 & 4 & 72 &-12.29 &   25.53  &   3.24  &30.20 & 23.00 &   1225\\ 
   16 & 0 & 72 & -8.37 &    8.37  &   0.0   & &       &    324\\ 
   12 & 4 & 72 &-10.63 &   20.63  &   0.0   & &       &    939\\ 
\hline
   12 & 8 & 76 &-12.29 &   40.05  &   7.76  &30.20 &  41.17 &   2212\\ 
   16 & 4 & 76 & -2.46 &   15.62  &   3.15  &20.72 &22.85 &    867\\ 
   14 & 6 & 76 & -4.90 &   19.90  &   0.0   & &       &    987\\ 
   16 & 4 & 76 & -6.23 &   16.23  &   0.0   & &       &    805\\ 
\hline
   12 & 12& 80 & {\bf -0.30} &   47.01  & 16.71&30.20 &  49.15 &   2792\\ 
   16 & 8 & 80 &  1.76 &   27.45  &   7.69&20.72  &  41.17 &   1733\\ 
   18 & 6 & 80 & -0.04 &   15.04  &   0.0 &  &        &    823\\ 
\hline
   12 & 16& 84 &  5.91 &   51.92  &  17.82  &30.20&  54.71 &   3271\\ 
   16 & 12& 84 & {\bf 13.47} &   33.20  &  16.67  & 20.72& 49.01 &   2240\\ 
   20 & 8 & 84 &  6.05 &   13.95  &   0.0&   &        &    840\\ 
\hline
   12 & 16& 84 &  3.02 &   51.25  &  22.26  &30.20&  54.05 &   3223\\ 
   20 & 8 & 84 &  2.05 &   13.95  &   0.0   & &       &    840\\ 
\hline
\end{tabular*}
\end{table}
To ascertain the stability of the estimates, all the calculations\tcn{,}
 done with $\zeta=0$, have been redone for $\zeta=3$. The examples
that follow are for $A=84$ which involves the largest magnitudes for
$hsp,\, l\epsilon_{gr}$ and $E_q$ and hence, presumably, the largest
uncertainties.

For $\zeta=0\to 3$ the energies $E=$5.91, 13.47 and 6.05 in
Table~\ref{tab:central} go to 4.03, 11.89 and 5.0 respectively,
leaving unchanged the qualitative conclusions that may be drawn.

The evolution of the monopole is another source of uncertainty: The
$\varepsilon_i$ and $\varepsilon_{gr}$ numbers are suggested by
GEMO~\cite{gemo} at the beginning of the region. As the $g$ filling
increases, the orbit will separate from its $ds$ partners and come
closer to the $r_3$ space. To simulate this effect, in the last two
lines of Table~\ref{tab:central} the single particle energies are
changed to the bracketed values in the caption. As a consequence the
energies at $E=5.91$ and 6.05 change to 3.02 and 2.05 respectively. Again
the qualitative conclusions are not affected.

These results for $^{84}$Mo are typical and illustrate two important points:
\begin{enumerate}
\item For prolate states \bet and \q are very unsensitive to
  monopole behavior and hence remain close to their theoretical
  quasi+pseudo SU3 maxima. In our example $Q_4$=54.71 and 54.05 against
  the 55.88 maximum.
\item Energies of prolate states are very sensitive to the single particle
  field $H_{sp}$. In our example a shift of some 4.5 MeV:
  $hsp=17.82\, vs$ 22.26 MeV. However, the relative positions of the
  states remain fairly stable.
\end{enumerate}
Examine now what conclusions can be drawn from Table~\ref{tab:central}.

\vspace{0.2cm}
\noindent
$\bm {^{72}}${\bf Kr} The only species where \bet are close for
prolate and oblate candidates. Probable coexistence.

\vspace{0.2cm}
\noindent
$\bm {^{76}}${\bf Sr} Single candidate. Coexistence ruled
out. Experimentally superb rotor with good $J(J+1)$ sequence. Perfect
agreement of Table~\ref{tab:central} with a recent measure: $B(E2,
2^+$$\rightarrow$0$^+$)=~2220(270)~e$^2$fm$^4$ \cite{be2sr76}.

\vspace{0.2cm}
\noindent
$\bm {^{80}}${\bf Zr} The lowest state is expected to gain some 4 MeV
because of triaxiality and the observed rotational spectrum seems to
guarantee \bet close to the prediction. However the very low lying
oblate state may blur the picture. Moreover the prolate 8p-8h (and a
10p-10h not shown) are also close and triaxial. Finally, the
frustrated doubly magic $N=Z=40$ is at\ldots 0. MeV. A very
interesting nucleus.

\vspace{0.2cm}
\noindent
$\bm {^{84}}${\bf Mo} Strong hint of coexistence, even triple
coexistence through gains due to triaxiality of the second prolate
candidate.

Except for $^{76}$Sr,
coexistence is expected in the other nuclei and will be examined in
Section~\ref{sec:ccp}.

\subsection{The $\bm{r3g}$ calculations} 
Calculations in the $(r_3g)^n$ spaces have been carried out for all
$n$. We concentrate on results for $A\ge 80$. In particular $^{80}$Zr
and $^{84}$Mo are mainly of interest in lending support to a basic
observation about oblate bands:

Contrary to prolate states that privilege maximizing the deformation,
the oblate bands give precedence to mixing that reduces it. As a
consequence our schematic estimates overestimate \q and \bet and
underestimate energies.
 
\vspace{0.2cm}
\noindent
$\bm {^{80}}${\bf Zr} {\bf and}  $\bm {^{84}}${\bf Mo}
\vspace{-12pt}
\begin{table}[h]
\caption{\label{tab:zr80} $^{80}$Zr. Results of the full (1.1$10^{10}$
  dimensional) $r_3g$ calculation with {\sc jun45} (E's in MeV, $Q$ in
  efm$^2$ and \bet in e$^2$fm$^4$)}
\begin{tabular*}{\linewidth}{@{\extracolsep{\fill}}|c|ccccc|}
\hline 
J&  E(2$^+$) & $Q_{s}$&$Q_{0s}$&B(E2)&$Q_{0t}^2$\\
\hline
2& 0.393 & 51&{-179}&642&{$(180)^2$} \\
\hline
\end{tabular*}
\end{table}
\vspace{-6pt}
In Table~\ref{tab:zr80} the extracted $Q_0\approx 180$ is definitely
lower than the 6p-6h number from Table~\ref{tab:gr3}, $Q_0\approx
(19+23+3)b^2\approx$ 203 \ef.  The wavefunctions have 22\% 4p-4h, 44\%
6p-6h and 28\% 8p-8h. Mixing with prolate states nearby may be at the
 origin of the reduction, as confirmed in Table~\ref{tab:mo84} for
$^{84}$Mo.
\vspace{-6pt}
\begin{table}[h]
\caption{\label{tab:mo84} $^{84}$Mo. Properties of the yrast band;
  experiment vs.  calculations in the $r_3g$ space with the {\sc jun45}
  interaction: to the left truncated up to 4 holes in $r_3$. To the
  right complete space. (E in MeV, $Q$ in \ef and $B(E2)$ in
  \eff}
\begin{tabular*}{\linewidth}{@{\extracolsep{\fill}}|c|ccccc|cccc|}
\hline 
J & Ex & Et & $-Q_{0s}$ &  B(E2) &$-Q_{0t}$&
Et & $-Q_{0s}$ &  B(E2) &$-Q_{0t}$\\ 
  \hline 
 0  & 0.0  & 0.0  &     &    &     &0.00 &    &    &     \\ 
 2  & 0.44 & 0.17 & 194 &762 & 196 &0.29 &189 &708 &188  \\ 
 4  & 1.12 & 0.56 & 190 &1081& 195 &0.84 &189 &1020 &189  \\ 
 6  & 2.01 & 1.15 & 184 &1179& 194 &1.60 &189 &1118 &189  \\ 
       \hline 
    \end{tabular*}
\end{table}

The ground state band is dominated now by the $r_3^{-4}g^{8}$
configuration. From Table~\ref{tab:gr3} $Q_0\approx
(21+20+3)b^2=44\times4.61\approx 203$ \ef not inconsistent with the
{\em truncated} calculations (left of the Table) that exhibit good
rotational features. Once the full space is incorporated (right of the
Table), the energies depart from the $J(J+1)$ sequence while the
quadrupole properties, still those of a rotor, have suffered an
erosion due to the inclusion of prolate states as suggested in
$^{80}$Zr.

\vspace{0.2cm}
\noindent
$\bm {^{88}}${\bf Ru} 

\begin{table}[t]
\caption{\label{tab:ru88exp} Properties of the yrast band of
  $^{88}$Ru; experiment {\em vs} calculations in the $r_3g$ space with
  the {\sc jun45} interaction (E in MeV, $Q$ in efm$^2$ and $B(E2)$ in
  e$^2$fm$^4$)}.
\begin{tabular*}{\linewidth}{@{\extracolsep{\fill}}|ccccc|}
\hline  
J & E(exp) & E(th) & -Q$_s$ &  $B(E2\downarrow)$(th) \\ [2pt]
  \hline 
 0$^+_1$  & 0.0   & 0.0   &      &         \\ [2pt]
  2$^+_1$ & 0.62  & 0.56  & 37 & 492  \\ [2pt]
 4$^+_1$  & 1.42  & 1.31  & 44 & 766   \\ [2pt]
 6$^+_1$  & 2.38  & 2.12  & 47 & 888       \\ [2pt]
  8$^+_1$ & 3.48  & 2.88  & 52 & 980       \\ [2pt]
       \hline 
    \end{tabular*}
\end{table}
In $^{88}$Ru we come at last to a genuine $r_3g$ nucleus. (Note that
for $A\ge 88$ most numerical results reported below duplicate those of
the Rutgers group~\cite{zamickRuPd}).  Table~\ref{tab:ru88exp}
corresponds to an yrast oblate band exhibiting 50\% $r_3^{-4} g^{12}$
oblate dominance.
Not obvious, since $g^{12}$ is now beyond midshell and the largest \q
is prolate.  However the oblate \q in $r_3^{-4}$ is sufficiently
strong to dominate but the prolate admixtures  distort and reduce the
original \q=-(18.66+20.24) and $Q_0\approx -182$ to $Q_{0s}\approx
-125$ and $Q_{0t}\approx -160$ in Table~\ref{tab:ru88exp}. It is seen
that in this nucleus the prolate-oblate competition within the $r_3g$
space is played up. $^{92}$Pd will bring further news.

\vspace{0.2cm}
\noindent
$\bm {^{92}}${\bf Pd}
   
The authors' interest in heavy N=Z nuclei was sparked by the first
measurement of the $^{92}$Pd spectrum, accompanied by an
interpretation that associated it to a condensate of $(g_{9/2}^2)$
neutron-proton pairs coupled to maximum
\mbox{$J=9$}~\cite{ceder:10,chong:11,vanI}. Which raised two issues: that
of possible coupling schemes in a $g^{12}$ space, and that of possible
dominance of this configuration. Table~\ref{tab:pd92}---which we
comment column by column---sums up sufficient information to resolve
both issues:

\vspace{-0.1cm}

\begin{table}[h]
\caption{Properties of $^{92}$Pd. Energies in
  MeV, $Q$ in efm$^2$ and $B(E2)$ in e$^2$fm$^4$. Detailed explanation
  in text\label{tab:pd92}}
\begin{tabular*}{\linewidth}{@{\extracolsep{\fill}}|c|cccccccccc|}
\hline
  1 &2  &3 &      4                   &5  &6  &7   &8& 9&10 & \\ 
  J &Ex&Et&con&$qq$& $\Omega$&$B(E2)$&$B(E2)_{r_3g}$&$Q_s$ &$Q_{s,r_3g}$&\\
\hline
  0 & 0.0    & 0.0  &   .00 & .00 & .99 & ---  &---  &---  &---    & \\
  2 & 0.874  & 0.84 &   .26 & .22 & .99 & 225  &304  & -28 &-3.63  & \\
  4 & 1.786  & 1.72 &   .58 & .62 & .99 & 316  &382  & -34 &-8.20  & \\
  6 & 2.563  & 2.52 &   .85 &1.20 & .98 & 340  &364  & -31 &-2.77  & \\
\hline
\end{tabular*}
\end{table}

\noindent
\tcn{1. J value}
 
\vspace*{.1cm}
\noindent
2. Experimental spectrum, in very good agreement with

\vspace*{.1cm}
\noindent
3. {\sc jun45} spectrum.

\vspace*{.1cm}
\noindent
4. Spectrum of the condensate defined by $-H_{\rm con}$ = P$_0$ +9P$_{9}$,  
where P$_0$ and P$_9$ are the pairing Hamiltonians for $J=0$ and 9.

\vspace*{.1cm}
\noindent
5. Spectrum of the quadrupole force scaled so as to have unit $J=9$
matrix element. Close to that of the condensate (within arbitrary
scaling factor) 

\vspace*{.1cm}
\noindent
6. Overlap, \mbox{$\Omega=\langle qq|{\rm con}\rangle^2$}, of the
wavefuntions indicating that the condensate and quadrupole coupling
schemes are identical. The use of P$_9$ should be understood as an
artifact to define a coupling scheme. As a Hamiltonian it is better
avoided.

\vspace*{.2cm}
\noindent
7, 8. Now for the second issue. A Hamiltonian $-H \approx .6qq+.4P_0$
yields $g^{12}$ energies that are close to the exact ones and $B(E2)$
that are very close to the pure $qq$ values in column 7, and not too
far from the exact ones in column 8. Which may encourage the idea of
$g^{12}$ dominance in spite of its smallish 30\% contribution to the
exact wavefunction. However, this idea is not supported by the
disparity of $Q_s$ in columns 9 and 10.

\vspace*{.1cm}
\noindent
9, 10 Spectroscopic $Q_s$ for $qq$ \tcn{(9)} and {\sc
  jun45}~\cite{jun45}(10).

\vspace*{.2cm}
\noindent
\tcn{The situation} is reminiscent of that of $f_{7/2}^n$ configurations
that yield apparently reasonable energetics and transition rates but
quadrupole moments of the wrong sign~\cite{pozu:81}. 

The pattern we started following at $^{80}$Zr---of oblate states
progressively eroded by prolate mixtures---now reaches its climax with
the Pyrrhic victory of prolate states practically cancelled by oblate
mixtures.

\vspace{0.2cm}
\noindent
$\bm {^{96}}${\bf Cd}  

For this nucleus, the calculations in the $r_3g$ and $g$ spaces with
{\sc jun45} give results that are much closer than in $^{92}$Pd, both
for the energies and for the $B(E2)$ properties and the discrepancies in
the spectroscopic quadrupole moments are gone except for the $6^+$
state.
\begin{table}[h] \caption{$^{96}$Cd. Energies in MeV, $Q$ in \ef and $B(E2)$ in \eff
  e$^2$fm$^4$\label{tab:cd96}}
\begin{tabular*}{\linewidth}{@{\extracolsep{\fill}}|rccccccrrr|}
\hline
   & &$\Delta E$ &  & &$B(E2)$ & & &$Q_s$&\\\hline
 $J^{\pi}$         &   $r_3g$  &  $g_{9/2}$ & $sdg$ &  $r_3g$ & $g_{9/2}$ & $sdg$ & $r_3g$ & $g_{9/2}$ & $sdg$  \\ \hline
0$^+$        &   0.0  &  0.0      &  0.0 &     &     &     &     &     &\\
2$^+$        &   0.90 &  0.96     & 0.77 & 152 & 154 & 327 & -19 & -23 & -37\\
4$^+$        &   1.91 &  2.10     &1.78  & 203 & 206 & 426 & -22 & -22 & -40\\
6$^+$        &   3.02 &  3.08     & 2.78 & 191 & 159 & 351 & -11 & -5  & -23\\
8$^+$        &   3.48 &  3.08     & 3.24 & 47  & 40  & 65  & 40  & 39  & 55\\
\hline
\end{tabular*}
\end{table}
We have collected some results in Table \ref{tab:cd96}, adding
those from the full $sdg$ space using the Nowacki-Sieja
interaction~\cite{nara:11} which describes the superallowed 
decay of $^{100}$Sn~\cite{nat_sn100} and the $B(E2)$ systematics of the
light Sn isotopes~\cite{guastalla:13}. The results for the energies,
$B(E2)$ and $Q$ values vary little between $g$ and $r_3g$ pointing to
$g$ dominance, not invalidated by the substantial quadrupole
coherence brought in by the full $sdg$ space calculation as it amounts
basically to an overall scaling. 

It is worth mentioning that the latter predicts a  16$^+$ isomer at 5.3~MeV.

\section{Case Studies, Comparisons and Perspectives}\label{sec:ccp}
The central region calls for some extra comments.
\subsection{Coexistence in $^{72}$Kr}\label{sec:kr72}

%\begin{table}[h]
%\caption{\label{tab:kr72exp} Properties of the yrast band of
%  $^{72}$Kr; experiment vs.  calculations in the $r_3g$ space with the
%  {\sc jun45} interaction
%  (see text) (E's in MeV, $Q$ in efm$^2$ and $B(E2)$ in e$^2$fm$^4$)}
%\begin{tabular*}{\linewidth}{@{\extracolsep{\fill}}|cccccc|}
%\hline 
%J & E(exp) & E(th)  & Q$_{0t}$ &  $B(E2\downarrow)$(th) &  $B(E2\downarrow)$(exp)\\ [2pt]
%  \hline 
% 0$^+_1$  & 0.0    & 0.0  &      &       &  \\ [2pt]
%  2$^+_1$  & 0.71 & 0.35  &207 & 850  & 999(129)\\ [2pt]
% 4$^+_1$  & 1.32  & 1.11  &206 &1210  & \\ [2pt]
% 6$^+_1$  & 2.11  & 2.27  &204 &1300    &   \\ [5pt]
% 8$^+_1$  & 3.11  & 3.77  &200 &1310  & \\ [5pt]
% 0$^+_2$  & 0.67  &       &      &          & \\ [2pt]
%  \hline
%    \end{tabular*}
%\end{table}
%
Exact $(r_3g)^{16}$ calculations with \jun  for $^{72}$Kr indicate
that---with respect to Table~\ref{tab:central}---the gap
\tcn{$r_3^{12}g^4-r_3^{16}$} is underestimated by about 2 MeV, while \bet is
overestimated by 10\%. The groud state band is a nice oblate rotor
with nearly constant $Q_{ot}\approx 205$\ef, good $J(J+1)$ sequence
with the $2^+$ at 350 keV---half the observed value---while the  $4^+$
at 1.1 MeV is close to the observed 1.32 MeV, while \bet=850\eff
against a measured 999(129)e$^2$fm$^4$\cite{be2kr72}. In this
reference it is argued that the ground band is oblate. A suggestion
that may gain some support from the shape of the Gamow Teller
$\beta^+$ decay strength function~\cite{sarri-gt}. Recent
measures~\cite{iwa} yield $B(E2:2_1\to 0_1)$=810(150)\eff (too small
to be prolate) and $B(E2:4_1\to 2_1)$=2720(550) \eff. (too large to be
oblate\tcn{)}. Even large for prolate in view of theoretical maximum of 2200
\eff. (Note that an analysis of Fig. 3 of~\cite{iwa} suggests that the
550 \eff error bar is underestimated).

%%          KIT
%$J=\quad ~~~ 2 ~~~~\quad 4 ~~~~~\quad 6$
%
%$Q_{0t}^2=[50.27,~~~~ 35 .19,~~~~~ 31.95~~~30.52]$\bet 
%
%$Q_{0s}=b^2*Q_{0s}/b^2[-3.5,\,-2.75,\, -2.5\, -2.11]Q_{spec}(J)$

Clearly some mixing is necessary and to achieve it we resort to the
space which is the largest we could treat and the smallest that could
cope with prolate states \ie $r_3^{16-t}(gd)^t$, tractable for $t\le
4$. The interaction chosen is {\sc r3gd} \ie  {\sc jun45} supplemented
by matrix elements involving the $d$ orbit from the {\sc lnps}
set~\cite{lnps}.

First two calculations at fixed $t=4$ were made. If the
single-particle energy $\varepsilon_d$ is set 1.76 MeV above
$\varepsilon_g$, the ground state band is solidly prolate. If the
splitting is incresed by 0.5 MeV the lowest $J=0^+$ and $2^+$ become
oblate, but the lowest $4^+$ is prolate and nearly degenerate with its
oblate counterpart. The two bands simply slide past, ignoring each
other. To achieve any mixing, extreme fine tuning is required.
\begin{table}[t]
\caption{\label{tab:kr72th} Properties of the yrast bands of $^{72}$Kr
  calculated in the $r_3gd$ space with the {\sc r3gd} interaction (see
  text) (E's in MeV, $Q$ in efm$^2$ and $B(E2)\equiv B(E2:J_i\to
  J_{fx})$ in e$^2$fm$^4$). Bottom, first line: measured values
  from~\cite{iwa} (error bars subject to caution as explained in
  text), second line: $t\le 4$ results boosted as explained in the
  text.}
\begin{tabular*}{\linewidth}{@{\extracolsep{\fill}}|ccccc|cccc|}
\hline 
&$t=4$&&\multicolumn{2}{c|}{$B(E2)$}&$t\le 4$&&\multicolumn{2}{c|}{$B(E2)$}\\  
\hline
$ J_i$&   $Ex$ &$Q_s$  &$J_{f1}$&$J_{f2}$& $Ex$ &$Q_s$ &$J_{f1}$&$J_{f2}$ \\
$ 0_1$&   0.0  &      &     &     &    0.00& 	&	  &          \\
$ 0_2$&   0.24&       &     &     &    0.30& 	&	  &          \\
$ 2_1$&   0.28& -65   &1089 &    6&    0.46& -54&   586& 372 \\
$ 2_2$&   0.56&  58   &   3 &  897&    0.66&  45&   103& 536 \\
$ 4_1$&   0.83& -77   &1509 &    1&    1.05& -75&  1387&  75  \\
$ 4_2$&   1.23&  69   &   0 & 1286&    1.43&  64&    36&1093 \\
\hline
    \end{tabular*}
\begin{tabular*}{\linewidth}{@{\extracolsep{\fill}}|l|}
$B(E2:2_1\to 0_1)$=810(150),~~~~~$B(E2:4_1\to 2_1)$=2720(550)~~\\
$t\le 4\times$1.4; $B(E2:2_1\to 0_1)$=740,~~~$B(E2:4_1\to 2_1)$=1750\\
\hline
    \end{tabular*}
\end{table}
Things change when configuration mixing is allowed. In
Table~\ref{tab:kr72th}, to the left, the result at fixed $t=4$ with
prolate ground state ($\varepsilon_d-\varepsilon_g=1.76$ MeV). The
choice is made to present the two bands in their pure form.  To the
right, the $t\le 4$ results show prolate dominance with strong ground
state mixing, using $\varepsilon_d-\varepsilon_g=2.26$ MeV which
yields oblate ground state at fixed $t=4$. While \bet is halved the
$B(E2: 4_1^+\to 2_1^+)$ changes by less than 10\%. To estimate the
effect of omitting the $s$ orbit we redo calcutations as in
Table~\ref{tab:central} $\varepsilon_d=2.$ MeV and no $s$ orbit, and
then add $\varepsilon_s=3.$ MeV. The rates are boosted 15\%, and a
further 10\% may come from $e_0=2.1$ as suggested near the end of
Section~\ref{sec:se68}, for a total of 26\%. At the bottom of
Table~\ref{tab:kr72th} the corresponding boosted values are compared
with the observed ones. Let us add two entries to the list of
calculations quoted by Iwasaki and coworkers~\cite{iwa}, which fall in
two groups.
 \begin{itemize}
\item Those that mix prolate and oblate states. They include
  Vampir~\cite{vampir00}---which produces thorough mixing as a
  reassessment of oblate dominance previously
  predicted~\cite{vampir}---and the relativistic mean field work of Fu
  {\em et al.}~\cite{fu}, close to the present results: strong
  $0^+_1\pm 0^+_2$ mixing and strong prolate dominance in $2^+_1$.
\item Those that predict oblate ground bands. They include
  Skyrme~\cite{benderbh,64-84} and Gogny~\cite{girod,krtomas} (beyond)
  mean-field approaches and a sophisticated form~\cite{sato} of the
  Kumar-Baranger model in two major oscillator shells~\cite{bk68}.
\end{itemize}
In next section~\ref{sec:mono} it will be explained why a
majority of calculations privilege the oblate solution.

{\sf Digression} Gamow-Teller strength calculated with the present
wavefunctions agrees nicely with observation.
\subsection{Monopole {\em vs} Single particle field}\label{sec:mono}
Most mean-field-based calculations hve single particle spectra in
which the $d$ orbit is some 5 MeV above the $g$ one (as
in~\cite[Fig.1]{zr80tomeg}) \ie some 2 MeV above the values in
Table~\ref{tab:central}. As emphasized in \cite{rmp} and section II $H_m$ is a strict
two body operator, but its action can be simulated by single particle
fields---provided it is understood that they vary as a function of the
orbital occupancies---the 'monopole drift', mostly due to the filling
of the largest $j$ orbit in a major shell~\cite{az576}. In the
problems studied here, the $r_3$ space can be viewed as frozen, but
the $gds$ orbits are subject to drift. Above $^{56}$Ni the $r_3$
orbits are nearly degenerate and the $gds$ ones are close to an
$l\cdot s$ sequence. There is no direct experimental evidence for the
position of the $ds$ orbits around $A=68$ but we can rely on the GEMO
program~\cite{gemo} which accounts for the particle or hole spectra on
all known double magics to within 200 keV and confirms the $l\cdot s$
behavior with the $d$ orbit 2-3 MeV {\tcn above} the $g$ one, which
upon filling comes closer to $r_3 $ and becomes detached from $ds$
which move up to join their $r_4$ partners to form a pseudo $LS$
sheme. 

In $^{72}$Kr, as elsewhere, the structure of the states is unsensitive
to monopole details but their energies are nor. Which explains why so
many calculations place the prolate state too high.

\subsection{Potential Energy Surfaces in $^{80}$Zr.}\label{sec:PESZr} 
In the comments to Table~\ref{tab:central} it was noted that to the
three states included for $^{80}$Zr one should add a 10p-10h,
$r_3^{14}(gds)^{10}$ prolate state at about 1 MeV with \bet$\approx$2100\eff,
and the $r_3^{24}$ closed shell at 0 MeV. As the three prolate states
are triaxial they will gain energy (of the order of 3-4 MeV according
to Table~\ref{tab:gds8-12}) and dominate the low lying spectrum.
In the thorough study of {\tcn Rodr\'{\i}guez} and Egido with a Gogny
force~\cite[Fig.3]{zr80tomeg} this is very much the case. The main
discrepancy with their work is in the positioning of the closed shell,
which in the potential energy surface~\cite[Fig.2]{zr80tomeg} comes
some 4 MeV below the deformed minima which we attribute to
underbinding of the latter due to the monopole effect described above.
\tcn{Other calculations for $^{80}$Zr include~\cite{vampir,64-84,srzrmo-zheng}.}

\vspace{-12pt}

\subsection{Coexistence in $^{84}$Mo}\label{sec:Mo} 
According to~\cite{vampir,64-84} the ground state of $^{84}$Mo is prolate,
and spherical respectively. Table~\ref{tab:central} suggests three
candidates:

I) A splendid axial rotor (all ``platforms'' filled in the ZRP
diagrams in Fig~\ref{fig:su3}), expected to have a 2$^+$ well below
the observed 440 keV.

II) A splendid oblate rotor (Table~\ref{tab:mo84}) whose  2$^+$ is
  way too low.

III) A triaxial rotor.  

No direct information is available but $^{82}$Zr, whose behavior is
likely to be similar, provides a hint. Collating data from~
\cite{zr82_1,zr82_2,zr82_3}: the ground state starts with $2^+$
at 407 keV and \bet$\approx$1900\eff, is interrupted by a definitely
smaller $B(E2:4^+\to 2^+)$ at 700-1200\eff and then resumes with a
fairly constant $Q_0$, albeit smaller than the one extracted from
\bet. \tcn{Our guess: prolate dominance in $^{84}$Mo quenched by
  mixing.}

\vspace{-12pt}

\subsection{Beyond intrinsic states}\label{sec:pp}

In mean-field studies, ``going beyond'' amounts to projecting and
moving in the $\beta-\gamma$ plane. Here, we do not have a potential
energy surface but a space of discrete intrinsic states: The numerous
local minima, revealed by Tables~\ref{tab:ex} and ~\ref{tab:gr3},
constitute a natural basis in which pairing will act as mixing agent.  We
do not know yet how to do the mixing, but we may call attention to a
way of dealing with each individual state: 

Diagonalize separately in the quasi and pseudo spaces and then
recouple in the product space. There is nothing new with this
``weak-coupling'' idea except one thing: the quadrupole force(s) used
in each space must be boosted as much as necessary to reproduce the
quadrupole moments dictated by the MZ calculations.
\section{Looking back and forward}
The operation of the quasi-pseudo SU3 tandem was shown to account for
the onset of rotational motion in the rare earths, involving the
$r_4hfp$ (proton) and $r_5igds$ (neutron) EEI spaces~\cite{zrpc}.  The
formal basis of this successful estimate was not clear at the
time. Now it can be ascribed to Nilsson-SU3 selfconsistency that puts
together the two classics in the field: the Bohr Mottelsson rotational
model~\cite{bm53} plus Elliott's quadrupole force and SU3
symmetry~\cite{su3}, via a reinterpretation of the Nilsson
model~\cite{nilsson}. Eq.~(\ref{hfin}) makes explicit the connections
\begin{gather}
H=H_{sp}-\frac{\hbar \omega \delta}{3}2q_0\equiv H_{sp}-\beta\hbar \omega \kappa
\frac{\langle 2q_0 \rangle}{{\cal N}^2} 2q_0\label{hfin}
\end{gather}
To the left, the Nilsson problem amounts to calculate single particle
energies in the presence of a deformation $\delta$. The constraints on
$\delta$ were left undefined and the earliest successful calculation
of quadrupole moments relied on volume conservation~\cite{mn59}. It
was much later that the Nilsson orbits could be associated to an
energy minimization, either via the Strutinsky~\cite{strut} or
Hartree-Fock-Bogoliubov (HFB)~\cite{bk68} methods.

To the right of Eq.~(\ref{hfin}) we have summed up the selfconsistent
formulation \tcn{---the MZ Eqs.~(\ref{hmql1},\ref{hmql2})---} with its
built-in constraint: the input \q must coincide with the output
\q. The emphasis is on the quadrupole moment, not the energy: Nilsson
orbits are sensitive to the central field while ``\q orbits'' \ie
\tcn{the ZRP diagrams in} Fig.~\ref{fig:su3} are nearly constant,
reflecting the underlying operation of pseudo-quasi SU3. The resulting
interpretive framework explains the appearance of ``closed shells''
(the $(\lambda 0)$ representations), the natural prolate dominance,
the importance of ``triaxiality'' (the $(\lambda \mu)$ representations)
and the \tcn{abrupt departure from the $pf$ regime beyond $A=68$}.

The open task is to put the energetics on firmer ground, and to go
beyond intrinsic states as hinted at the end of the previous
section~\ref{sec:pp}.  The challenge is to keep the approach simple,
or, at least, computationally doable.

\vspace{6pt}

%\vspace{-12pt} 

\begin{acknowledgments} 
This paper is dedicated to Joaqu\'in Retamosa, co-inventor of
quasi-SU3, on the first anniversary of his untimely disappearance.

%Roberto Liotta contributed encouragement and useful remarks.

We thank F. Recchia for an illuminating discussion on the recent
important $B(E2)$ measurements of $^{72}$Kr~\cite{iwa}.

\tcn{This work is partly supported by Spanish  grants FPA2011-29854 from MICINN 
and SEV-2012-0249 from MINECO, Centro de Excelencia Severo Ochoa Programme.}

%\vspace{12pt} 
{\bf Notes on authorship}

A.P.Z designed, wrote and contributed to all sections

All authors contributed to sections V A and V C.

A. P. contribute to section III

S. M. L. contributed to section II 

\end{acknowledgments}

\end{document}